\documentclass[useAMS,usenatbib]{mn2e}
\usepackage[pdftex]{graphicx}
\newcommand{\Swift}{\textit{Swift}}

\newcommand{\ip}{\textit{i$^{\prime}$}}
\newcommand{\zp}{\textit{z$^{\prime}$}}

\newcommand{\rp}{\textit{r$^{\prime}$}}
\newcommand{\J}{\textit{J}}
\newcommand{\Hm}{\textit{H}}
\newcommand{\Ks}{\textit{K$_{s}$}}

\def\lesssim{\mathrel{\hbox{\rlap{\hbox{\lower3pt\hbox{$\sim$}}}\hbox{\raise2pt\hbox{$<$}}}}}
\def\gtrsim{\mathrel{\hbox{\rlap{\hbox{\lower3pt\hbox{$\sim$}}}\hbox{\raise2pt\hbox{$>$}}}}}
\newcommand\ion[2]{#1$\;${\small\rmfamily\@Roman{#2}}\relax}

\title{Evidence for Dust Destruction from the Early-time Colour Change of GRB\,120119A}

\author[Morgan et al.]{Adam N. Morgan$^{1,*}$,
             Daniel A. Perley$^{2}$,
             S. Bradley. Cenko$^{1}$,
             Joshua S. Bloom$^{1}$,
\newauthor
             Antonino Cucchiara$^{3}$,
             Joseph W. Richards$^{1,4}$,
             Alexei V. Filippenko$^{1}$,
             Joshua B. Haislip$^{6}$,
\newauthor
            Aaron LaCluyze$^{6}$,
            Alessandra Corsi$^{7}$,
            Andrea Melandri$^{8}$,
            Bethany E. Cobb$^{7}$,
\newauthor
            Andreja Gomboc$^{9,10}$,
            Assaf Horesh$^{2}$,
            Berian James$^{1,11}$,
            Weidong Li$^{1,12}$,
\newauthor
            Carole G. Mundell$^{8}$,
            Daniel E. Reichart$^{6}$,
            Iain Steele$^{8}$. \\
$^{1}$Department of Astronomy,
  University of California, Berkeley, CA 94720-3411, USA \\
$^{*}$\texttt{amorgan@astro.berkeley.edu} \\
$^{2}$Department of Astronomy, California Institute of Technology, Pasadena, CA 91125, USA.  \\
$^{3}$Department of Astronomy and Astrophysics, UCO/Lick Observatory, University of California, Santa Cruz, CA 95064, USA \\
$^{4}$Department of Statistics, University of California, Berkeley, CA 94720-3860, USA \\
$^{5}$Department of Physics and Astronomy, University of North Carolina, Chapel Hill, NC 27599-3255, USA \\
$^{6}$INAF -- Brera Astronomical Observatory, Merate (LC), Italy \\
$^{7}$Department of Physics, The George Washington University, Washington, DC 20052 USA \\
$^{8}$Astrophysics Research Institute, Liverpool John Moores University, Birkenhead CH41 1LD, UK \\
$^{9}$Faculty of Mathematics and Physics, University of Ljubljana, Slovenia\\
$^{10}$Centre of Excellence SPACE-SI, Slovenia \\
$^{11}$Dark Cosmology Centre, Copenhagen ¯, Denmark \\
$^{12}$Deceased 12 December 2011\\
}

\begin{document}





\pagerange{\pageref{firstpage}--\pageref{lastpage}} \pubyear{2013}

\maketitle

\label{firstpage}

\begin{abstract}
We present broadband observations and analysis of \Swift\ gamma-ray burst (GRB) 120119A. 
Our early-time afterglow detections began under 15\,s after the burst in the host frame (redshift 
$z = 1.73$), and they yield constraints on the burst energetics and local environment.  Late-time 
afterglow observations of the burst show evidence for a moderate column of  dust 
($A_V\approx1.1$ mag) similar to, but statistically distinct from, dust seen along Small Magellanic 
Cloud sightlines. Deep late-time observations reveal a dusty, rapidly star-forming host galaxy.  
Most notably, our early-time observations exhibit a significant red-to-blue colour change in the first 
$\sim 200$\,s after the trigger at levels heretofore unseen in GRB afterglows.  
This colour change, which is coincident with the final phases of the prompt emission, is a hallmark 
prediction of the photodestruction of dust in GRB afterglows.  
We test whether dust-destruction signatures are significantly distinct from other sources of colour 
change, namely a change in the intrinsic spectral index $\beta$. 
We find that a time-varying power-law spectrum alone cannot adequately describe the observed 
colour change, and allowing for dust destruction (via a time-varying $A_V$) significantly improves
 the fit.  While not definitively ruling out other possibilities, this event provides the best support
  yet for the direct detection of dust destruction in the local environment of a GRB.   
\end{abstract}

\begin{keywords}
gamma-ray burst: individual (GRB\,120119A) -- dust, extinction
\end{keywords}

\section{Introduction}
\label{sec:intro}
More than seven years after the launch of \Swift~ \citep{gehrels04}, a rich 
collection of optical light curves of gamma-ray bursts (GRBs) has been 
amassed. While global similarities in light-curve behaviors are becoming 
well established \citep[e.g., ][]{kann10a,kann11a}, thorough studies of individual 
bursts nevertheless continue to yield important insights into the details of the 
explosions and their environments. In particular, there remain relatively few 
examples of very early-time optical light curves when the prompt phase of the 
high-energy emission is still ongoing  
\citep[for some examples, see ][]{panaitescu08a,melandri08a,cenko09a,rykoff09a,oates09a} 
due to the small but significant delays in relaying positional information to the 
ground, and the nonzero response times of even the fastest robotic telescopes.

Furthermore, the majority of the earliest-time light curves are of a single colour 
(often unfiltered), which can provide only limited information on the evolution
 of the GRB emission and interactions with its surrounding environment.  
 Contemporaneous, multi-colour observations are needed to identify and 
 characterize any colour change that may occur in the early afterglow.

Strong colour change is expected to occur for at least some events owing to
the photodestruction of dust in the nearby environment of GRBs at early times 
\citep{waxman00,draine02}. Given the association of massive-star progenitors 
with long-duration GRBs, it is natural to expect GRBs to explode in dusty environments.  
However, only a small fraction ($\sim 25\%$) exhibit evidence for significant 
($A_V > 1$ mag) dust obscuration in late-time optical/ultraviolet (UV) spectral 
energy distributions (SEDs) \citep{covino13a}.  One potential explanation of this 
apparent discrepancy is the photodestruction of dust in GRB environments by 
prompt high-energy emission.  

So far, however, unambiguous evidence for or against dust destruction in early 
GRB afterglows has been varied and inconclusive.  As the bulk of the destruction 
is likely produced by bright X-rays during the prompt emission, optical observations 
contemporaneous with the high-energy radiation are needed in order to observe 
time variability in the SED \citep[e.g.,][]{fruchter01,perna03}.   

Indirect evidence for or against dust destruction has been explored in observations 
of a few individual GRBs. In GRB 030418, an optical rise following a dearth of  emission 
 seen in images beginning $\sim 200$\,s after the start of the burst ($> 1$\,min after the 
prompt phase ended, and thus after dust destruction would have occurred) is interpreted 
as absorption of optical photons by dust inside a massive stellar wind medium \citep{rykoff04}. 
In this interpretation, complete destruction of the local dust did not occur, as some must 
have persisted to produce the attenuation. However, other models can be invoked to 
explain this early rise and subsequent decay that is now commonly seen in early-time 
optical afterglows (see, e.g., \S\ref{sec:earlyrise}).

In the case of GRB 061126, \citet{perley08a} and \citet{gomboc08a} explore the possibility 
that grey dust could explain the observed deficit of optical flux to X-rays at late times.  The 
optical dimness of this event was interpreted as optical absorption without the expected 
reddening by normal dust observed in the local universe.  While the presence of grey 
dust was not conclusively seen, a possible interpretation of its existence would be the 
sublimation of smaller grains in the local environment due to early photoionization of 
dust by the prompt X-ray emission.  However, a significant change of $A_V$ was not 
reported, so the grey dust might have been pre-existing.

In order to unambiguously identify the signatures of dust destruction, simultaneous, 
multi-colour imaging of the afterglow  (especially during the high-intensity high-energy emission) is necessary.
With well-sampled, multi-colour optical/near-infrared (NIR) observations beginning less 
than 1\,min after the burst trigger, GRB\,120119A offers one of the best cases yet to study the nature of 
an early-time GRB afterglow.   
The afterglow exhibits an appreciable red-to-blue colour change coincident with the end 
of the prompt high-energy emission, making this event an excellent candidate to test 
for the signatures of dust destruction.
However, careful modelling of the SED as a function of time is needed to disambiguate 
between changes in the dust-absorption properties and other sources of colour change, 
such as a change in the intrinsic spectral index $\beta$.

In this study, we present observations and analysis of the broadband afterglow of 
GRB\,120119A.  Details of the observations and data reduction are given in \S\ref{sec:obs}.  
Next, we present modelling and analysis, including general properties of the afterglow 
behavior in \S\ref{sec:lc}, modelling of the late-time extinction profile in \S\ref{sec:latesed}, 
details of the construction of early-time SEDs in \S\ref{sec:sedvstime}, modelling of the 
early-time colour change in \S\ref{sec:colorchange}, constraints on the origin of the early 
emission in \S\ref{sec:earlyrise}, and details of the host-galaxy properties in \S\ref{sec:host}.
To conclude, we discuss the implications of the results in \S\ref{sec:discussion}.
Throughout the paper we adopt the specific flux convention of $F(\nu,t) \propto \nu^\beta t^\alpha$.  
All quoted uncertainties are $1\sigma$ unless stated otherwise, and dates and times are given in UT.

\section{Observations}
\label{sec:obs}

\subsection{{\it Swift} Observations}
\label{sec:swiftobs}

GRB\,120119A triggered the Burst Alert Telescope \citep[BAT;][]{barthelmy05}
onboard \Swift~\citep{gehrels04} at 04:04:30.21.  \Swift~slewed immediately to the source and began observations with the X-Ray Telescope \citep[XRT;][]{burrows05} and UltraViolet/Optical Telescope \citep[UVOT;][]{roming05} at 53.3\,s and 61\,s after the trigger, respectively.  

We processed the BAT data using the formalism of \citet{butler07b}.  The BAT data show a duration of $T_{90} = 70 \pm 4$\,s and a total fluence (15--350 keV) of $S = (2.3 \pm 0.1) \times 10^{-5}$\,erg\,cm$^{-2}$.
The XRT data were reduced using the pipeline described by \citet{butler07a}, and they were corrected for Galactic neutral hydrogen (assuming the standard associated absorption from other elements) using the maps of \citet{kalberla05}. The XRT bandwidth covers the energies 0.2--10\,keV; all flux values quoted herein are converted to an effective energy of 1\,keV.

The beginning of the XRT observations marginally overlaps with the end of the observed BAT emission; this initial steep decline of X-ray emission ($\alpha = -2.71 \pm 0.09$) is consistent with an extrapolation of the tail of the BAT emission to lower energies, as has been seen in many previous bursts \citep[e.g.,][]{obrien06}.

The optical afterglow was clearly detected with the White filter onboard the UVOT in the initial finding-chart exposure taken $\sim 140$\,s after the burst ($m_{\rm white} = 19.5 \pm 0.1$ mag), and was again marginally detected in the $B$ band $\sim 1800$\,s after the trigger \citep[$B = 19.4 \pm 0.3$ mag;][]{chester12a}.  It was undetected in all other filters, including the first $U$-band exposure $\sim 290$\,s after the trigger ($m_{U} > 18.7$ mag).

\subsection{PROMPT Observations}

Observations of GRB\,120119A were taken by the Skynet robotic telescope network using 5 PROMPT telescopes at Cerro Tololo Inter-American Observatory (CTIO) in Chile. Observations began at 04:05:08, 38\,s after the burst.  Observations in $I,R,V,B,$ and open filters were taken beginning at different starting times, but nearly simultaneous multi-colour observations were performed when possible.  Unfortunately, the images take in the $V$ band suffered from a previously unknown detector issue, and thus were not used in the analysis. The exposure of each observation increases with time since the GRB, with a minimum of 5\,s to a maximum of 80\,s in the $I,R,B,$ and open filters.
Stacking is done manually as necessary to increase the signal-to-noise ratio (S/N), with an aim of obtaining photometric uncertainties of approximately $\pm 0.1$ mag.

Photometry was performed using a custom pipeline written in C and Python, based upon IRAF\footnote{IRAF is distributed by the National Optical
Astronomy Observatory, which is operated by the Association of Universities for Research in
Astronomy (AURA), Inc., under cooperative agreement with the US National Science
Foundation (NSF).} aperture photometry.  Photometric calibration was performed using a selection of 7 SDSS stars. For filters not in SDSS, colour transformations were performed using the prescription of \citet{jester05}. The results are shown in Table \ref{tab:GRB120119Aphot}, where magnitudes are in the Vega system.

\subsection{PAIRITEL Observations}
\label{sec:ptel}
The robotic Peters Automatic Infrared Imaging Telescope \citep[PAIRITEL;][]{bloom06} began automatic observations of GRB\,120119A at 04:05:23, 53\,s after the BAT trigger.  PAIRITEL consists of the 1.3\,m Peters Telescope at Mt. Hopkins, AZ, which was formerly used for the Two Micron All Sky Survey \citep[2MASS;][]{skrutskie06} but was subsequently refurbished with the southern 2MASS camera. PAIRITEL uses two dichroics to image in the NIR \J, \Hm, and \Ks~ filters simultaneously every 7.8\,s.  Three images are taken at each dither position and then median combined into 23.4\,s ``triplestacks.''  Images are then resampled to $1''$ pixel$^{-1}$ from its native $2''$ pixel scale and coadded using SWarp \citep{bertin02}.

Aperture photometry was performed using custom Python software, utilizing Source Extractor \citep[SExtractor;][]{bertin96} as a back end. 
The optimal aperture of $5.5\arcsec$ diameter was determined by minimizing the absolute error relative to 2MASS magnitudes of our calibration stars.
Calibration was performed by redetermining the zeropoint for each image individually by comparison to 2MASS magnitudes with the calibration stars. The resulting statistical uncertainty in the zeropoint is negligible relative to other sources of error. Additional, systematic sources of error are addressed in detail by \citet{perley10}; we use a similar procedure here to determine the total uncertainty of each point. The results are shown in Table \ref{tab:GRB120119Aphot}, where magnitudes are in the Vega system.

\subsection{KAIT Observations}
\label{sec:kait}
The Katzman Automatic Imaging Telescope \citep[KAIT;][]{filippenko01a} began observations of GRB\,120119A at 04:06:33, 2.05 min after the BAT trigger. KAIT is a 0.76\,m telescope located at Lick Observatory dedicated to discovering and observing supernovae and other transients; it has been autonomously responding to GRB triggers since 2002 \citep{li03b}.

KAIT began observations by cycling through 20\,s exposures in $V$, $I$, and unfiltered images, later switching to alternating $I$ and unfiltered exposures, and finally (when the source was faint) just unfiltered images.  Automatic bias subtraction and flat-fielding is performed at the telescope.  Coaddition with SWarp is then performed as necessary to obtain significant detections of the afterglow.  

Aperture photometry with a $2\arcsec$ radius was performed on the coadded images via a custom IDL wrapper based on the GSFC IDL Astronomy User's library {\tt aper} routine.
Calibration was done relative to field stars from the Sloan Digital Sky Survey Data
Release 8 \citep[SDSS DR8;][]{aaa+11a}, where magnitudes were converted into $V,R,I$ using the transformation equations of Lupton (2005)\footnote{http://www.sdss.org/DR7/algorithms/sdssUBVRITransform.html\#Lupton2005}.  Unfiltered observations were calibrated relative to $R$-band magnitudes following the procedure of \citet{li03a}.  The results are shown in Table \ref{tab:GRB120119Aphot}, where magnitudes are in the Vega system.

\subsection{Liverpool Observations}
The 2\,m Liverpool Telescope\footnote{http://telescope.livjm.ac.uk/} (LT) robotically responded to the BAT trigger under its automatic GRB follow-up program \citep{guidorzi06} and began observations $\sim 2.6$\,min after the burst.  The first $\sim 10$\,min of observations were obtained with the RINGO2 polarimeter, which were coadded into a total of seven 80\,s frames and then calibrated against the SDSS \rp~ filter. Subsequent observations were acquired by alternating SDSS \rp\ip\zp~ filters from 14.5\,min to 53\,min after the burst, with a pre-determined sequence of increasing exposures. Differential photometry was performed with respect to 5 SDSS field stars with the Graphical Astronomy and Image Analysis Tool (GAIA).  The results are shown in Table \ref{tab:GRB120119Aphot}, where magnitudes are in the AB system.

\subsection{SMARTS Observations}
\label{sec:smarts}
Beginning at 04:41:41 ($\sim 0.64$\,hr after the burst), data were obtained using the ANDICAM (A Novel Dual Imaging CAMera) instrument mounted on the 1.3\,m
telescope at CTIO\footnote{http://www.astronomy.ohio-state.edu/ANDICAM}.
This telescope is operated as part of the Small and Moderate Aperture Research
Telescope System (SMARTS) consortium\footnote{http://www.astro.yale.edu/smarts}.
The ANDICAM detector consists of a dual-channel camera that allows for simultaneous optical and IR imaging.

During each epoch, multiple dithered images were obtained with an image cadence designed to ensure 
that the final combined frames in each filter are referenced to the same time of mid-exposure. Thus, afterglow measurements are obtained at a single reference time for all filters, without any need for temporal  extrapolation.  Standard IRAF data reduction was performed on these images,
including
bias subtraction, 
flat-fielding, and sky subtraction. The images were then aligned and averaged to produce a single master frame for each epoch. During the first epoch, total summed exposure times equaled 180\,s in \textit{BRIJK} and 120\,s in \textit{HV}.  For all other epochs, the total summed exposure times amounted to 15\,min in \textit{IV} and 12\,min in \textit{JK}.

The afterglow brightness was measured using seeing-matched, relative aperture photometry, with the relative magnitude
of the afterglow determined by comparison with a set of field stars.  The relative magnitudes were converted
to apparent magnitudes by comparison with the Rubin 149 standard star \citep{L92} in the optical and
with 2MASS stars in the IR.

\subsection{P60 Observations}

We began observations of the afterglow of GRB\,120119A with the automated 
Palomar 60\,inch (1.5\,m) telescope (P60; \citealt{cfm+06}) beginning at
7:33 on 2012 January 19 (3.48\,hr after the \textit{Swift}-BAT
trigger).  Images were obtained in the Sloan $g^{\prime}$, $r^{\prime}$,
and $i^{\prime}$ filters, and individual frames were automatically reduced
using our custom IRAF software pipeline.  To increase the S/N,
individual frames were astrometrically aligned using the Scamp software
package and coadded using SWarp \citep{bertin02}.

We used aperture photometry to
extract the flux of the afterglow from these coadded frames with the
aperture radius roughly matched to the 
FWHM of the point-spread function (PSF). Aperture magnitudes were then
calibrated relative to field sources from SDSS DR8.  Imaging continued on subsequent nights with the
P60 until the afterglow was below our detection threshold.
 The results are shown in Table \ref{tab:GRB120119Aphot}, where magnitudes are in the AB system.

\subsection{Gemini-S spectroscopy}
\label{sec:redshift}
On January 19.20, 53\,min after the BAT trigger,
we utilized our rapid Target-of-Opportunity program
(GS-2011B-Q-9, P.I. Cucchiara) to observe the optical
afterglow with GMOS-S on the Gemini South telescope. We obtained two spectra of 900\,s
each, with the R400 grating (corresponding to
a resolving power of $R \approx 1200$ at 6000\,\AA) and a $1''$ slit, 
spanning the 4000--8000\,\AA\ wavelength range. 
Flat-field and Cu-Ar lamp calibration files were obtained 
immediately after the target observation.

The data were reduced using the {\tt Gemini} and 
{\tt GMOS} packages available under the IRAF environment. 
Cosmic-ray rejection was performed using the
{\tt lacos-spec} routine \citep{vandokkum01}.
The extracted 1-dimensional spectra were combined and normalized
using the {\tt long\_combspec}, {\tt x\_continuum} and {\tt x\_nrmspec}
routine available under the XIDL package.

\begin{figure*}
  \includegraphics[width=18cm]{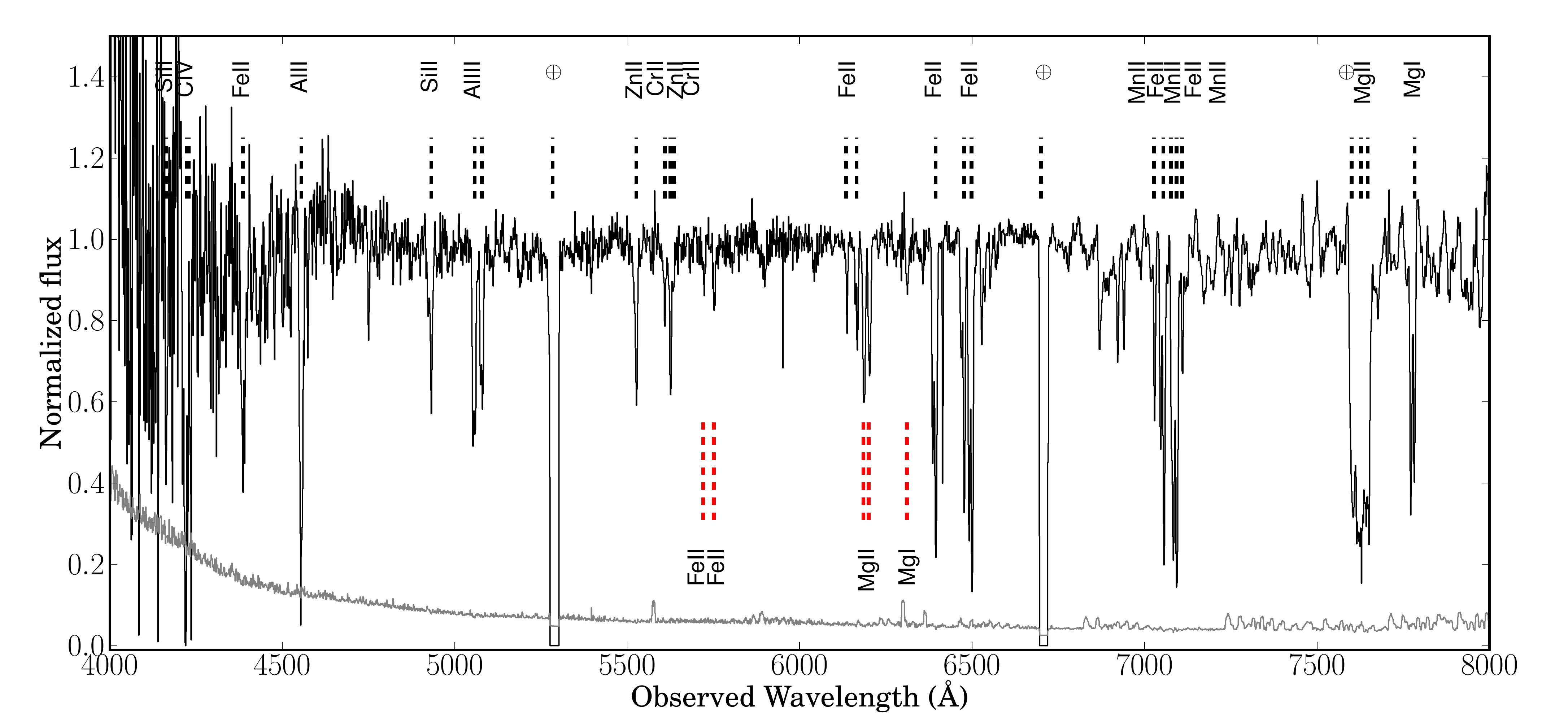}
  \caption{Normalized, combined spectrum of GRB\,120119A obtained with GMOS-South. Two systems of absorption features
are clearly identified. Indicated in black are the host-galaxy ($z=1.728$) metal lines, while in red
we see a strong Mg~II absorber at $z=1.212$, based on Mg~II and Fe transitions. At the bottom, we show the noise spectrum.  
 }
\label{fig:spectrum}
\end{figure*}

The final result is shown in Figure \ref{fig:spectrum}. The spectrum reveals several metal absorption features,
which are associated which a host galaxy at redshift $z=1.728 \pm 0.05$. In addition, 
a strong Mg~II system at lower redshift ($z=1.212$) is present  \citep[as is common in GRB afterglow spectra; e.g.,][]{vergani09,cucchiara09,prochter06}.

\subsection{Gemini-N Observations}

On January 21.28 (2.1 days after the burst) we observed the field of GRB\,120119A with the 
Gemini Multi-Object Spectrograph
\citep[GMOS;][]{hook04} and  Near InfraRed Imager and Spectrometer \citep[NIRI;][]{hodapp03}
as part of our rapid ToO program GN-2011B-Q-34 (P.I. Tanvir) in order 
to continue monitoring the late-time behavior of the afterglow.
We performed a series of short ($\sim 3$ min) exposures in the optical $g^{\prime}$,
$r^{\prime}$, and $i^{\prime}$ bands. These data were 
obtained with a dithered random pattern around the GRB location
 to improve the subsequent reductions. Data were flat-fielded and coadded
using the \texttt{Gemini-GMOS} tasks under the IRAF environment.
The final coadded images consist of a total of 26 min ( $i^{\prime}$) and
20 min ($r^{\prime}$ and $g^{\prime}$), with a scale of 0.14$\arcsec$ pixel$^{-1}$.

The IR observations consist of 20 dithered images of 80\,s each  ($4 \times 20$\,s exposures) 
in the $J$ band, 18 dithered images of 80\,s each ($16 \times 5$\,s exposures) in the $H$ band,  and 19 dithered images of 128\,s each ($16 \times 8$\,s exposures) in the $K$ band.
Each sequence was obtained with dithering patterns similar to those for the GMOS data.
Reduction, including cosmic-ray rejection, flat-fielding, and coaddition,
was performed using the {\tt NIRI} package.
The afterglow is detected in all of the final coadded images.

We obtained two additional NIRI $K$-band imaging epochs of the field of GRB\,120119A using our standard ToO program (GN-2011B-Q-10, P.I. Fox) on January 25 ($\sim 6$ days post trigger) and April 2 ($\sim 74$ days post trigger), with total integration times of 29\,min and 34\,min, respectively.  Seeing conditions were exceptional during both integrations ($0.35 \arcsec$).  Images were reduced and combined following the procedures above.  In both of these coadds we detect a faint, marginally extended ($\sim 0.3\arcsec$), nonfading ($< 0.15$ mag at 95\% confidence) source underlying the GRB position; we suggest that it is the GRB host galaxy.

\subsection{Keck Observations}  

We imaged the location of GRB\,120119A with the Low Resolution Imaging Spectrometer \citep[LRIS;][]{oke95} on the Keck-I 10, telescope at three different epochs.  The first epoch was carried out on January 26 between 10:42 and 11:07 (7.3 days post-trigger) with the $g$, $R$, and $I$ filters.  While seeing conditions for most of that night were generally poor, these data were taken within a brief window of good seeing ($\sim 0.7\arcsec$).  An object is well detected at the afterglow position at a magnitude only slightly fainter than in the GMOS imaging several days earlier.   

A second epoch of imaging was taken on February 21 (33 days post-trigger), but seeing conditions were poor ($1.7\arcsec$) and the position of the GRB is blended with the neighboring galaxy $3\arcsec$ to the south.  Nevertheless, a source is still clearly present at the GRB location.  Photometry of this object was complicated by the poor seeing and blending, but we do not see any clear evidence of fading from the previous LRIS epoch.  

A final epoch of imaging was performed on 2012 December 11 (0.89\,yr after the burst) under relatively good seeing conditions ($\sim 1.0 \arcsec$).  Images were taken with the $B$, $R$, and RG850 (roughly $z'$) filters.  The host galaxy is clearly detected in all three bands. 
                     
\subsection{{\it Hubble Space Telescope} Observations}

The position around GRB\,120119A was observed in the NIR with the Wide Field Camera 3 (WFC3) on the {\it Hubble Space Telescope (HST)} as part of our Cycle 20 program (GO-12949, PI D. Perley) to investigate the host galaxies of dust-obscured GRBs.  Observations were obtained in the F125W (wide $J$) and F160W (wide $H$) filters on 2012 October 28 ($\sim 283$ days post trigger).  The exposure time totaled 1209\,s in each filter, taken in both cases at three dithered positions.

\begin{figure}
  \includegraphics[width=9cm]{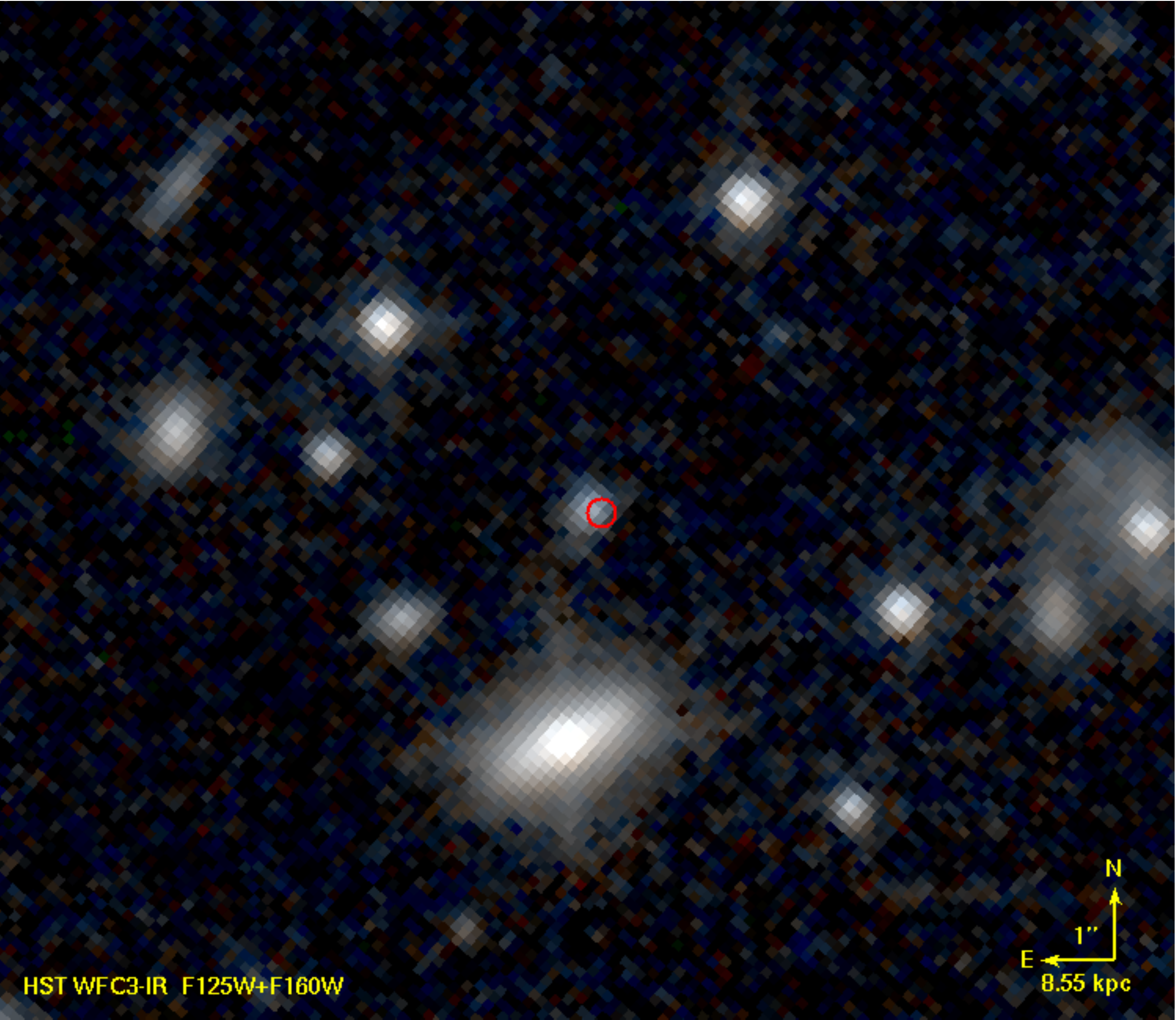}
  \caption{False-colour NIR {\it HST} imaging of the field of GRB\,120119A using the F125W and F160W filters of WFC3-IR, taken 9 months after the GRB.  The host galaxy appears as a compact source centred slightly east of the optical position (red circle) with some extension in the NW and SSE directions.  Given the redshift of the host, the blue colour suggests substantial line contribution to the $J$ band and a very large specific star-formation rate.  There are no obvious signs of ongoing interaction at this resolution.   The limiting magnitudes of the image are $F125W=25.6$ and $F160W=25.4$ (AB).
 }
\label{fig:hostimage}
\end{figure}

We downloaded the reduced observations from the Hubble Legacy Archive.  The host-galaxy candidate from our Keck and Gemini observations is clearly extended (Fig. \ref{fig:hostimage}).  It is easily detected (S/N $\approx 30$) in both frames, and it shows a morphology consistent with a bright, compact core ($r<0.3\arcsec$) surrounded by a disk of lower surface brightness ($r \approx 0.8\arcsec$).  There is no clear evidence of tidal features or ongoing interaction.  

Photometry of the host galaxy was performed using a custom IDL aperture photometry routine using a circular aperture of $1.0\arcsec$ radius centred on the host position.  We used the zeropoints in the WFC3 handbook and aperture corrections of 0.07 mag (F125W) and 0.09 mag (F160W) measured from stars elsewhere in the images.  The corresponding AB magnitudes of the host galaxy are $F125W = 23.13 \pm 0.04$ mag and $F160W = 23.34 \pm 0.07$ mag.

\subsection{CARMA Millimeter Observations}
\label{sec:carma}
We observed the location of GRB\,120119A with the Combined Array for Research in Millimeter-wave Astronomy (CARMA) for 2.2\,hr on 2012 January 19, beginning at 11:47:42. The observation were undertaken at a frequency of 95\,GHz with a bandwidth of 8\,GHz, while the telescope was in the C configuration. We used the compact source J0744-064 as a phase calibrator and observed Mars for flux calibration. The data were reduced using the MIRIAD software\footnote{http://bima.astro.umd.edu/miriad/}.  
The mid-epoch of the CARMA observation, $\sim 8.9$\,hr 
since the burst, is close to the last optical/NIR observations that yielded a detection (at $t \approx 5.6$\,hr
since the burst; see Fig. \ref{fig:lightcurve}). The $3\sigma$ upper limit derived from the CARMA nondetection is 0.99\,mJy at 95\,GHz.

\section{Analysis}
\label{sec:analysis}

\subsection{Optical/NIR Light Curve}
\label{sec:lc}

\begin{figure}
  \includegraphics[width=9cm]{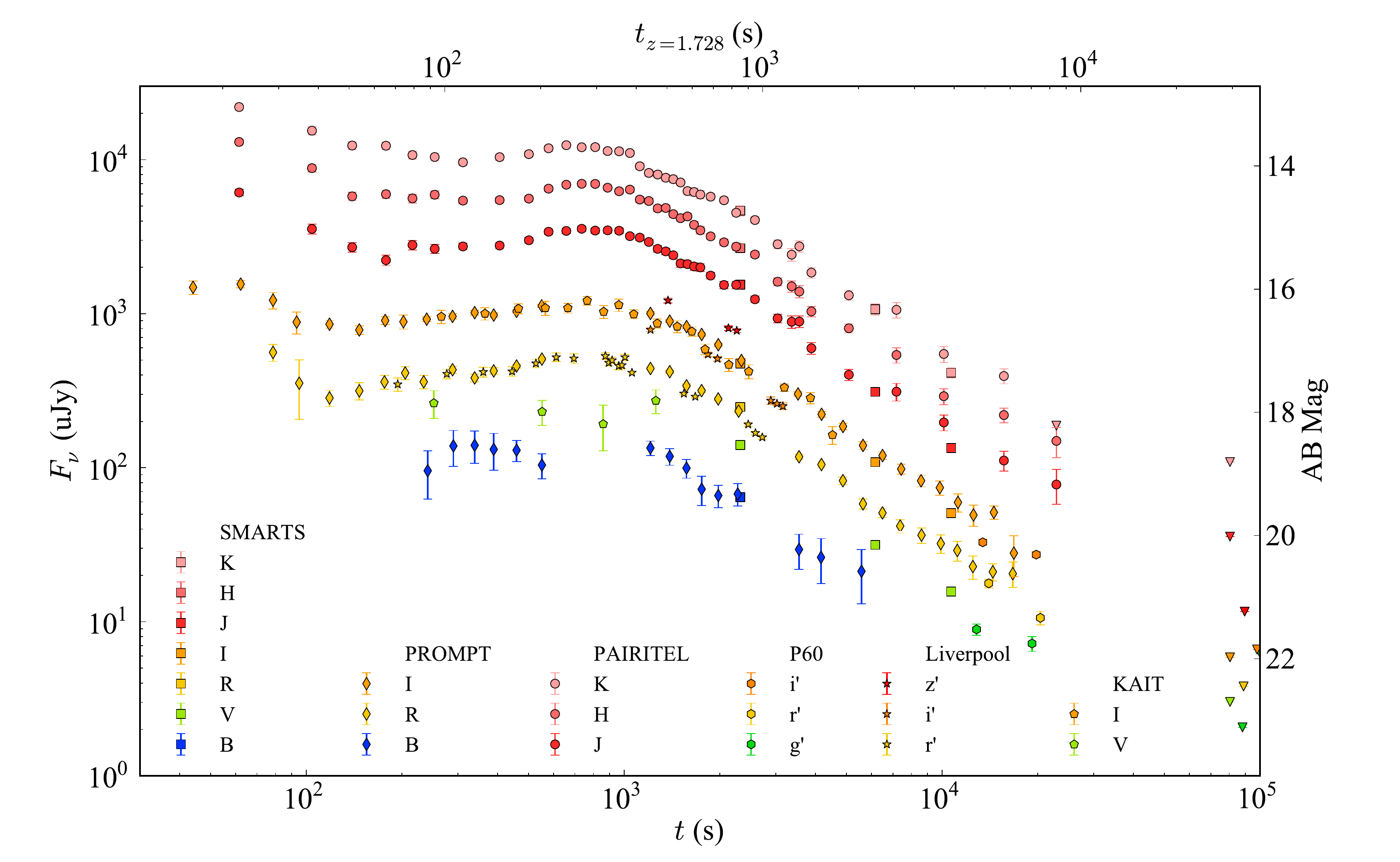}
  \caption{First-day optical/NIR light curve of GRB\,120119A. A significant red-to-blue colour change is observed during the first 200\,s after the trigger.  An achromatic rise is seen around 800\,s after the trigger, followed by some mild undulations before settling to a single power-law decay with index $\alpha_{\rm oir} = -1.30 \pm 0.01$. Unfiltered data are not included in this plot. The upper and lower time-axis scales refer to the rest frame and the observed frame, respectively. }
\label{fig:lightcurve}
\end{figure}

\begin{figure}
  \includegraphics[width=9cm]{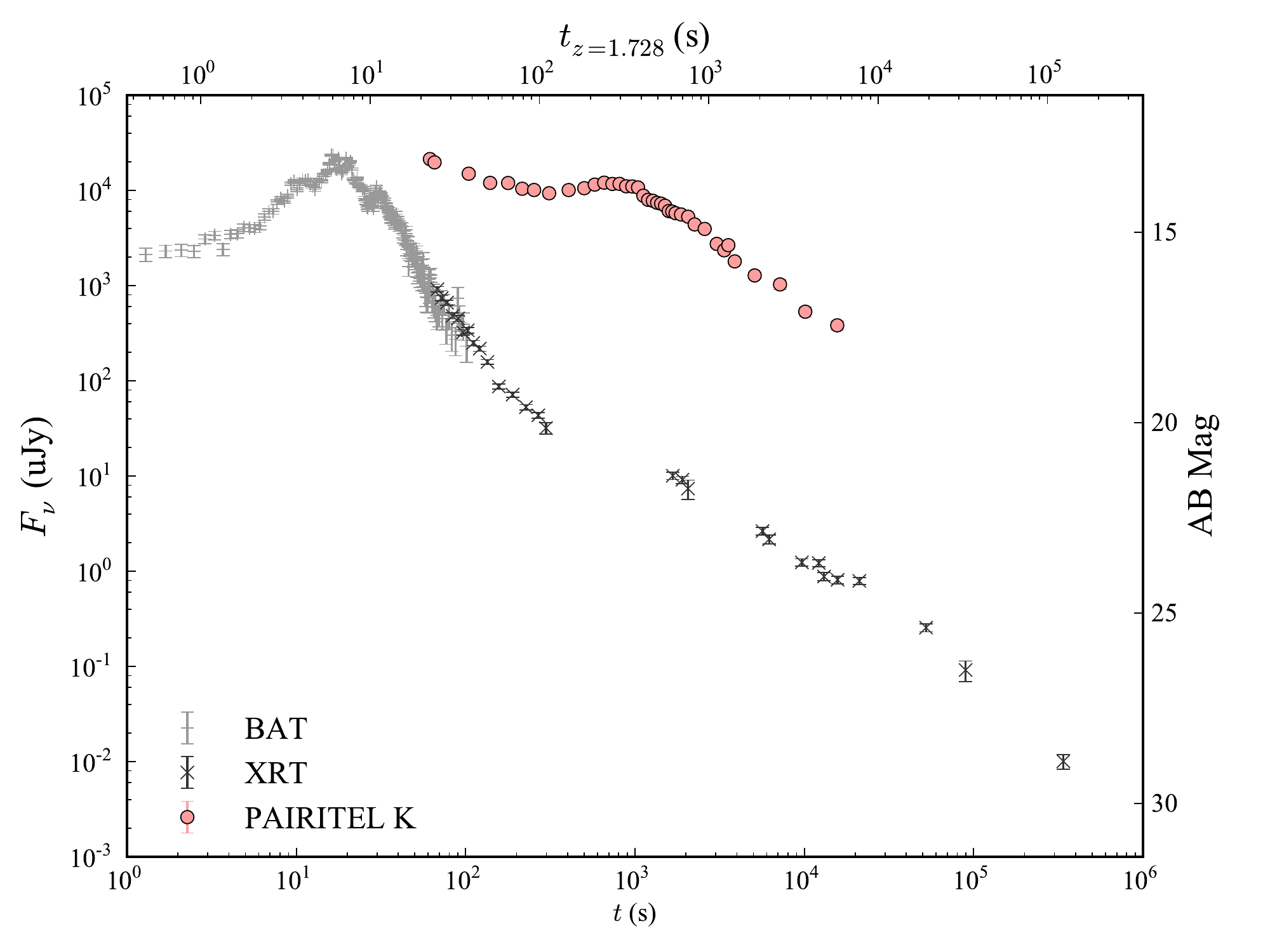}
  \caption{XRT light curve highlighting the overlap between the initial high-energy X-ray emission and the beginning of the optical observations during which significant colour change is seen for the first 200\,s.  The BAT data have been scaled up to match the XRT and are consistent with an extrapolation of the tail of the prompt BAT emission to lower energies. The upper and lower time-axis scales refer to the rest frame and the observed frame, respectively. 
 }
\label{fig:xrtlightcurve}
\end{figure}

A complete list of our afterglow photometry (uncorrected for Galactic extinction) is presented in Table \ref{tab:GRB120119Aphot}. All flux values used for modelling in this work have been corrected for the expected Galactic extinction of $E(B-V) = 0.093$ mag along the line of sight using the dust maps of \citet{schlegel98} and correcting for the $\sim14\%$ recalibration of these maps reported by \citet{schlafly11a}.
 The first day of optical/NIR photometry is plotted in Figure \ref{fig:lightcurve},
and some features are easily apparent.  First, a significant red-to-blue colour change is observed during the first 200\,s after the trigger.  We note that the bulk of this colour evolution occurs coincident with the end of the prompt high-energy emission (see Fig. \ref{fig:xrtlightcurve}). We explore the details and implications of this colour change in \S\ref{sec:colorchange}.

After the significant colour change ceased, the afterglow is seen to achromatically rise, peaking at roughly 800\,s (observer frame).  Minor oscillations are seen in some features during this peak, though coverage is limited and detailed modelling of these features is not presented here. We will explore the implications of the optical rise in \S\ref{sec:earlyrise}.

The XRT afterglow (plotted in Fig. \ref{fig:xrtlightcurve}) is well fit by a series of power-law decays. The afterglow is initially caught in a rapid decay 
which is consistent with a lower-energy extension of the tail end of the prompt gamma-ray emission seen by the BAT. The light curve then slows to a shallower decay of $\alpha_{X}=-1.27 \pm 0.02$ by the onset of the second epoch of observations. To compare this afterglow with that at lower energies (Fig. \ref{fig:xrtlightcurve}), we fit the optical/NIR light curve from the time of the onset of the second epoch of XRT observations until the end of the early-time optical observations ($\sim 0.4$ to 5.6\,hr after the burst).  While a single power-law component appears to dominate the light curve at this time, an additional excess of emission is seen at early times.  We fit the light curve using a combination of a single power law plus a \citet{beuermann99} component \citep[see][for a description of our light-curve fitting code]{perley08b}. The power-law component has a decay index $\alpha_{\rm oir} = -1.30 \pm 0.01$, consistent with the XRT decay at this time, indicating that the flux originates from the same synchrotron spectral component.  The minor rising component had a fixed rising index of $\alpha_{1,a} = 1.0,$ and a best-fit decay of $\alpha_{1,b} = 3.9 \pm 0.2$.

\subsection{Late-Time SED and Extinction Profile}
\label{sec:latesed}

\begin{figure*}  
  \begin{minipage}[b]{0.45\linewidth}
  \centering
  \includegraphics[width=9cm]{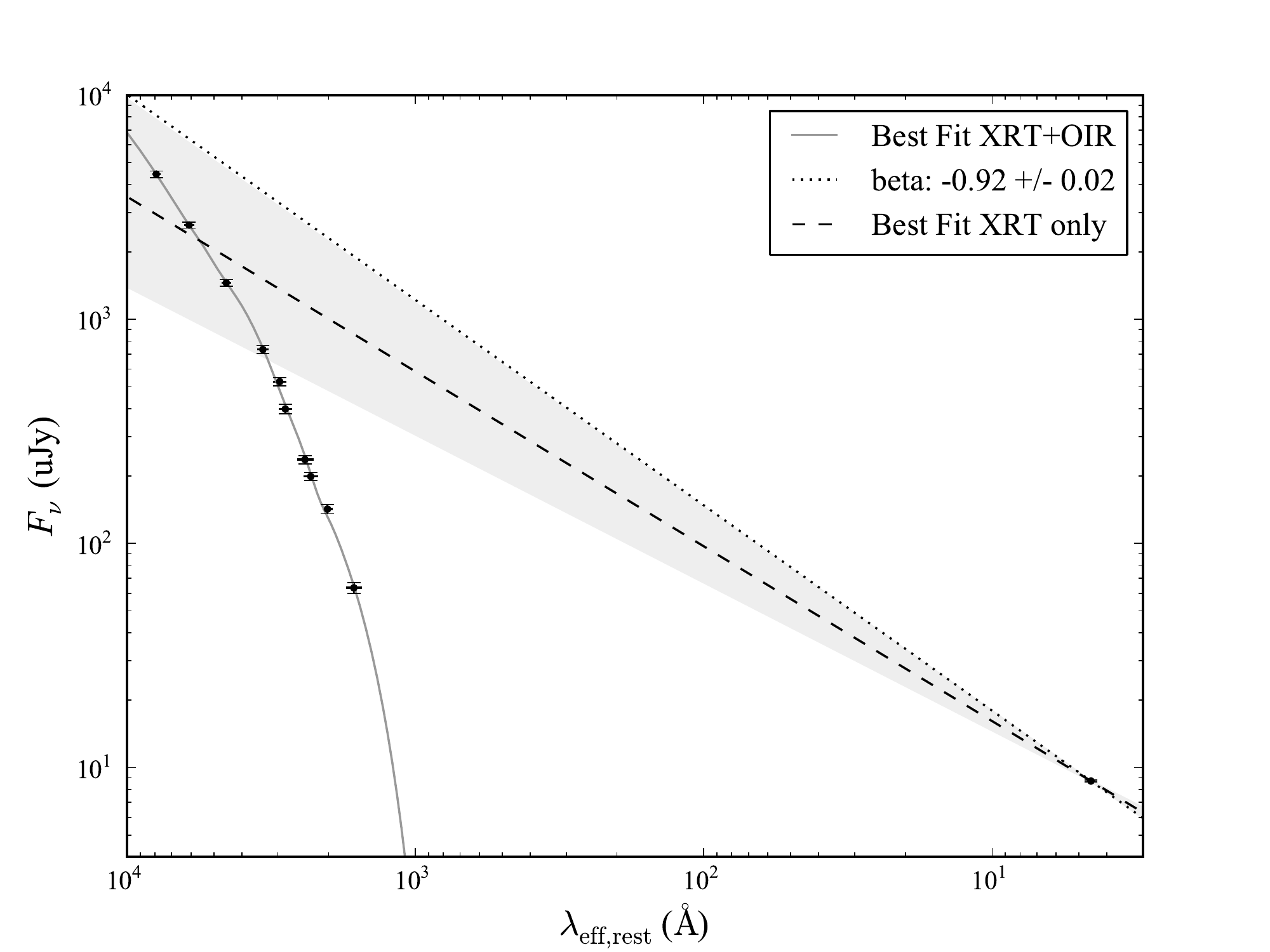}
  \end{minipage}
  \hspace{0.5cm}
  \begin{minipage}[b]{0.45\linewidth}
  \centering
  \includegraphics[width=9cm]{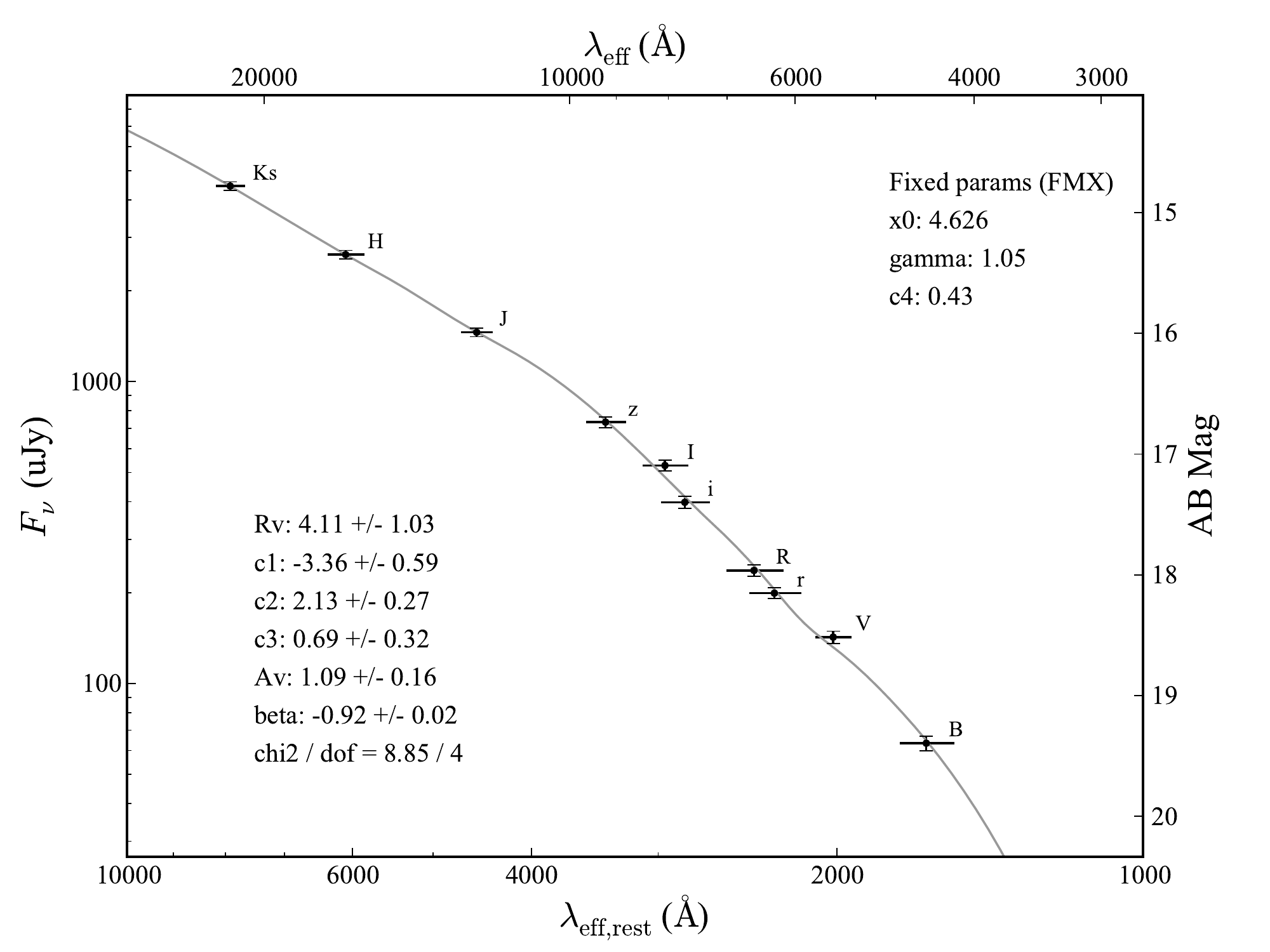}
  \end{minipage}
  \caption{Late-time SED of GRB\,120119A 
inferred from our light-curve model extrapolated to  38.7\,min after the burst (Table \ref{tab:SED}).
Values were corrected for the host redshift ($z=1.728$) and Galactic extinction ($A_V=0.288$ mag).  \textit{Left panel:} The dashed line and light grey cone represent the best-fit value and uncertainty (respectively) of $\beta$ inferred from the XRT spectrum alone using the products of \citet{butler07a}: $\beta_{\rm xrt} = 0.78 \pm 0.12$. The solid grey curve shows the best-fit dust model (FMX; $\chi^2$/dof = 8.9/4), and the dotted line is the $\beta$ inferred from that fit.  
\textit{Right panel:} Zoomed in SED highlighting the Optical/NIR points.  The horizontal bars at each point show the FWHM of the corresponding filter. The solid grey curve shows the best-fit dust model (FMX). 
 }
\label{fig:latesed}
\end{figure*}

At the time of the 
onset of the second epoch of XRT observations $\sim 20$\,min after the burst,
the complex evolution of the light curve ceased and given way to a simple power-law decay as the primary emission component (\S\ref{sec:lc}).  At this late stage, no evidence for further significant colour change is seen.  
To construct an SED from which to model the late-time dust extinction, we extracted the 
flux values in each band from our light-curve fit (\S\ref{sec:lc}) to the time 
of the first SMARTS observation 38.7\,min post burst. 
The resultant $1\sigma$ uncertainties on the fit parameters were multiplied by 
$\sqrt{(\chi^2/\rm{ dof})}$ for each corresponding filter to 
weight the uncertainty in each colour by an estimate of the individual 
light-curve fit qualities (which is particularly important for the $z$ and $V$ bands 
because of the relatively small number observations in these filters).
Estimated systematic uncertainties of 0.03 mag in the NIR and 0.04 mag in the optical 
were then added in quadrature. 
The resultant SED is given in Table \ref{tab:SED} and plotted in Figure \ref{fig:latesed}.

\begin{table}
  \centering
  \begin{minipage}{5.5cm}
  \caption{GRB 120119A Afterglow SED}
  \begin{tabular}{lcc}
  \hline
Filter & Flux\footnote{Afterglow fluxes were extrapolated to $t=38.7$\,min after the burst using our light-curve model.  Values have not been corrected for Galactic extinction.} & $1\sigma$ Uncertainty \\
& ($\mu$Jy) & ($\mu$Jy) \\
  \hline
$B $ &		$45.0 $ & $2.5$ \\
$V$ &      $109.9$ & $5.0$ \\
$r'$ &     $160.6$ & $6.6$ \\
$R$ &      $193.7$ & $8.2$ \\
$i'$ &     $340.6$ & $15.9$ \\
$I$ &      $456.1$ & $19.2$ \\
$z'$ &     $654.0$ & $26.2$ \\
$J$ &      $1358.6$ & $44.9$ \\
$H$ &      $2522.7$ & $80.8$ \\
$K$ &      $4312.1$ & $144.5$ \\

  \hline
  \end{tabular}

 \label{tab:SED}
  \end{minipage}
\end{table}

The SED clearly shows a significant dust component.  We fit various extinction laws to the SED using a custom Python implementation of the \citet{fitzpatrick99} parameterization based upon the GSFC IDL Astronomy User's library.  Average parameter values were extracted from \citet{gordon03a} for the Small Magellanic Cloud (SMC) and \citet{misselt99a} for the Large Magellanic Cloud (LMC).  Flux values and colours were corrected for the observed redshift ($z=1.728$; \S \ref{sec:redshift}) and Galactic extinction ($A_V=0.288$ mag, assuming $R_V=3.1$ for the Milky Way, MW), and the intrinsic spectral index $\beta$ was allowed to be free in all fits.  

Of the three standard local dust extinction laws (MW, LMC, SMC), the SED is by far best fit by an SMC curve ($\chi^2$/dof = 23.5/7), using the average value of the ratio of total to selective extinction $R_V=2.74$ from \citet{gordon03a}. Fits with $R_V$ as a free parameter were attempted, but lacking UV detections this parameter could not be well constrained.  For SMC dust, the best-fit values for extinction and spectral index are $A_V=0.62 \pm 0.06$ mag and $\beta = -1.39 \pm 0.11$.  

We can further constrain the value of $\beta$ by including the interpolated, unabsorbed XRT flux in the SED fit ($E_{\rm xrt,eff} = 1$\,keV; see \S\ref{sec:swiftobs}).
We note that the resultant best-fit value of $\beta$ from the optical/NIR data alone is somewhat large for this stage in the evolution and is inconsistent with the XRT observations under the assumption of a synchrotron-emission origin \citep[e.g.,][]{sari98}.  As the X-ray and optical/NIR afterglows are fading at approximately the same rate at the SED extrapolation time ($\alpha_{X}=-1.27 \pm 0.02$, $\alpha_{\rm oir} = -1.30 \pm 0.01$; see \S\ref{sec:lc}), we can assume they originate from the same component of the synchrotron spectrum.  With the inclusion of the XRT data in the SMC dust fit, we obtain best-fit values of $A_V=0.88 \pm 0.01$ mag and $\beta = -0.89 \pm 0.01$. 

That the SMC curve gives the best fit among the standard dust laws is consistent with the findings of previous studies \citep[e.g.,][]{schady12a,covino13a}. However, 
the goodness of fit is still poor ($\chi^2$/dof = 44.0/8) given the quality of the data.  
We thus attempted a more general fit using the full parameterization of \citet{fitzpatrick99}, which is described by six parameters: $c_1$ and $c_2$ are (respectively) the intercept and slope of the linear part of the UV component in $E(\lambda-V)$, $c_3$ is the strength of the 2175~\AA\ bump, $c_4$ is the strength of the rise
 in the FUV, $x_0$ is the centroid of the 2175~\AA\ bump in inverse microns ($x_0 \approx 1/0.2175$), and
 $\gamma$ is the width of that feature.

The values of $x_0$ and $\gamma$ are not seen to vary widely among dust sight lines within the Local Group, and thus we opt to keep these parameters fixed at $x_0=4.626$ and $\gamma=1.05$. 
Furthermore, the $c_4$ parameter which gives the far-UV extinction curvature at wavelengths $\lambda < 1700$ \AA\ is poorly constrained due to the lack of sufficiently blue filters in our data. As such, we fixed this parameter to the average value of $c_4 \approx 0.43$ found from the sample of \citet{schady12a}.  
All other parameters were allowed to vary freely.
The resultant fits are shown in Table \ref{tab:fmfits}.  While the parameter values without the XRT data included in the fit are poorly constrained, the best-fit dust law including the X-ray flux (henceforth FMX) shows a marked improvement over the other extinction laws ($\chi^2$/dof = 8.85/4; Table \ref{tab:extfits}). The best-fit FMX curve 
is shown in Figure \ref{fig:latesed} and yields values of $A_V=1.09 \pm 0.16$ mag and $\beta = -0.92 \pm 0.02$.

The best-fit FMX curve indicates that a weak 2175 \AA\ bump may be present ($c_3 = 0.69 \pm 0.32$), which at $z=1.728$ would lie nearly coincident with our $R$ band observations.  This feature has been more securely detected in several other GRBs, such as GRB 070802 \citep{kruhler08,eliasdottir09}, GRB 080603A \citep{guidorzi11a}, GRB 080605 \citep{zafar12a}, GRB 080607 \citep{perley11}, and GRB 080805 \citep{zafar12a}.

\begin{table}
  \centering
  \begin{minipage}{80mm}
  \caption{Results of Extinction Fits}
  \begin{tabular}{lcccc}
  \hline
Dust & +XRT? & $\beta$ & $A_V$ & $\chi^2$ $/$ dof\\
Model & &  & (mag) & \\
  \hline
LMC & N & $ -2.05 \pm 0.11$ & $ 0.37 \pm 0.07$ & 135.8 / 7 \\
LMC & Y & $ -0.92 \pm 0.00$ & $ 1.14 \pm 0.01$ & 230.1 / 8 \\
LMC2 & N & $ -1.03 \pm 0.16$ & $ 1.09 \pm 0.11$ & 69.5 / 7 \\
LMC2 & Y & $ -0.92 \pm 0.00$ & $ 1.16 \pm 0.01$ & 70.0 / 8 \\
MW & N & $ -2.50 \pm 0.09$ & $ 0.07 \pm 0.06$ & 160.5 / 7 \\
MW & Y & $ -0.94 \pm 0.00$ & $ 1.26 \pm 0.02$ & 464.6 / 8 \\
SMC & N & $ -1.39 \pm 0.11$ & $ 0.62 \pm 0.06$ & 23.5 / 7 \\
SMC & Y & $ -0.89 \pm 0.00$ & $ 0.88 \pm 0.01$ & 44.0 / 8 \\
FM & N & $ -1.04 \pm 3.27$ & $ 0.97 \pm 3.26$ & 8.9 / 3 \\
FM & Y & $ -0.92 \pm 0.02$ & $ 1.09 \pm 0.16$ & 8.9 / 4 \\

  \hline
  \end{tabular}

 \label{tab:extfits}
  \end{minipage}
\end{table}

\begin{table*}
  \centering
  \begin{minipage}{16.5cm}
  \caption{Best-Fit \citet{fitzpatrick99} Dust Parameters for GRB\,120119A}
  \begin{tabular}{lllllllll}
  \hline
Dust & $A_{V}$ & $\beta$ & $R_V$ & $c_1$ & $c_2$ & $c_3$ & $c_4$ & $\chi^2$ $/$ dof \\
Model & (mag) & (s) & (mag) & & & & & \\
  \hline
FM &  $0.97 \pm 3.26$ & $ -1.04 \pm 3.27 $ & $ 4.13 \pm 1.32 $ & $-3.76 \pm 12.17 $ & $ 2.29 \pm 4.91 $ & $ 0.75  \pm 1.82 $ & $ 0.43 {\rm (fixed)} $ & 8.9 / 3 \\
FMX & $1.09 \pm 0.16$ & $ -0.92 \pm 0.02 $ & $ 4.11 \pm 1.03 $ & $-3.36 \pm 0.59 $& $ 2.13 \pm 0.27$ & $ 0.69 \pm 0.32$ & $ 0.43 {\rm (fixed)} $& 8.9/4 \\

  \hline
  \end{tabular}
 
 \label{tab:fmfits}
  \end{minipage}
\end{table*}

The closure relations of standard afterglow theory \citep[e.g.,][]{granot02a,piran05a} can be used to infer the temporal decay index $\alpha$ and spectral index $\beta$ under various conditions through their relations with the electron spectral index $p$, allowing for a consistency check of our derived values for these two parameters. 
Under the assumption of an adiabatic expansion of the blast wave
into a homogeneous external medium ($n = \rm{constant}$) in the slow-cooling regime, there are two possibilities for the spectral index $\beta$ in the temporally decaying part of the light curve: $\beta = -(p-1)/2$ for $\nu_m < \nu < \nu_c$, and  $\beta = -p/2$ for $\nu_c < \nu$.
The two possible values for $\alpha$ in these regimes are $\alpha = -3(p-1)/4$
and $\alpha = -3(p-2)/4,$ respectively, leading to the closure relations $\beta = 2\alpha/3$ for $\nu$ between the peak frequency and the cooling break, and $\beta = 2(\alpha-1)/3$ for $\nu > \nu_c$. Adopting $\alpha \approx -1.30$ at the SED extrapolation time, we derive $\beta \approx -0.87$ for the $\nu_m < \nu < \nu_c$ closure relation, consistent with our more secure determination of $\beta = -0.92 \pm 0.02$ from the SED fit.  In this interpretation, the cooling break has not yet passed through our bandpasses and the electron energy distribution index is $p \approx 2.7$.

We can further check our derived value of $A_V$ by comparing it to the neutral hydrogen absorption excess as derived from X-ray data.  Using the photon-counting mode (PC) data from the XRT, we measure $N_{\rm H} = 1.03^{+0.46}_{-0.39} \times 10^{22}\, \rm{cm}^{-2}$.
\citet{watson12a} found that in addition to the general trend of a high $A_V$ requiring large values of X-ray absorption \citep{schady10a}, there is also a redshift trend indicating a dearth of low-$z$ events with low $A_V$ and high $N_{\rm H}$. This was expanded recently by \citet{covino13a}. While there is significant scatter in this relation, our derived values of X-ray absorption and optical extinction are in general agreement with the $N_{\rm H}/A_V$ trend of events at similar redshifts.

\subsection{Construction of Early-time SEDs}
\label{sec:sedvstime}

Our early-time observations are from a variety of telescopes and filters, at irregular times, and often rather sparsely sampled.  In order to model colour change as a function of time, we built up a series of SEDs at different temporal epochs using the PAIRITEL, PROMPT, Liverpool, and SMARTS photometry.  PROMPT clear-filter observations were not included in the SED constructions due to complications in accurate colour corrections with a changing sky background, and approximate redundancy with the equally well-sampled $R$ filter. 

One option for exploring colour change as a function of time is to identify epochs where data through multiple filters were obtained nearly simultaneously. We explored this option, but coincident temporal alignment of images was only achievable six times within the first 20\,min after the burst, leaving most of the data unused.

To circumvent this problem, interpolation was performed to create denser temporal sampling.  Because the PAIRITEL observations had the highest time resolution, these were chosen for interpolation. We use a nonparametric light-curve estimate to interpolate the GRB brightness (and measurement errors) in epochs of incomplete photometric coverage.  By using a flexible, nonparametric interpolation model, we avoid assuming an overly restrictive GRB template model and allow the data to determine the appropriate smooth light-curve shape.

To perform interpolation, we fit a natural cubic regression spline \citep[][]{ruppert2003semiparametric, wasserman2006all} separately to the data from each of the three PAIRITEL bandpasses.
We utilize regression splines because they are particularly adept at estimating smooth functions that may have complicated behavior such as periods of rapid decline and allow a straightforward estimate of both the brightness and model uncertainty of the brightness at each interpolation epoch (see, e.g., \citealt{richards12a} for usage in the context of SN light curves).
Using a cubic spline model, the estimate of the magnitude in photometric band $b$ at time $t$ is
\begin{equation}
\label{eqn:spline}
\widehat{m}_{b}(t) = \sum_{j=1}^{N + 4} \widehat{\beta}_b(t) B_j(t),
\end{equation}
where $N$ is the number of spline knots, $B_j$ is the $j^{\rm th}$ natural cubic
spline basis, and the $\widehat{\beta}_b(t)$ are estimates of the spline coefficients that are found by minimizing the weighted least-squares 
statistic of the spline-model magnitudes against the observed magnitudes $m_b(t)$ with weights $(1/\sigma_b(t))^2$.  Here, $\sigma_b(t)$ is the observational error in the measurement of $m_b(t)$.

When fitting a regression spline, one must choose the quantity, $N$, and the locations of the knots, which correspond to points of discontinuity of the third derivative of the interpolated spline function. In this work, we follow convention and place the knots uniformly over the observed time points, which have been converted to logarithmic space.  To choose $N$, we pick the value that minimizes the generalized cross-validation (GCV) score, which balances the bias and variance of the fitted interpolation function with an explicit penalty to avoid overfitting to the observed
data \citep[see, e.g.,][]{richards12a}.  The GCV criterion is defined as
\begin{equation}
\label{eqn:gcv}
{\rm GCV}_{b}(N) = \frac{1}{n_b}\sum_{k=1}^{n_b}\left(\frac{m_b(t_k) -\widehat{m}_b(t_k)}{\sigma_b(t_k)(1-N/n_b)}\right)^2,
\end{equation}
where $\widehat{m}_b(t_k)$ is the fitted value, at time $t_k$, of a spline with $N$ knots,
computed using Equation~(\ref{eqn:spline}); $t_1, t_2, ... t_{n_b}$ is the grid of epochs of observation in
the light curve for photometric band $b$; and $n_b$ is the number of observed epochs of the light curve.
In this formulation, the observational uncertainties in the measured magnitudes, $\sigma_b(t_k)$,
are used  to compute both the interpolated light-curve magnitude estimates, $\widehat{m}_b(t)$,
and the model uncertainty in those estimates.  
Using the GCV criterion, we search for the optimal number of spline knots over the range 1--50, finding the optimal value to be 14 for each of the $J$ and $H$ bands, and 8 for $K_s$.

\begin{figure}
  \includegraphics[width=9cm]{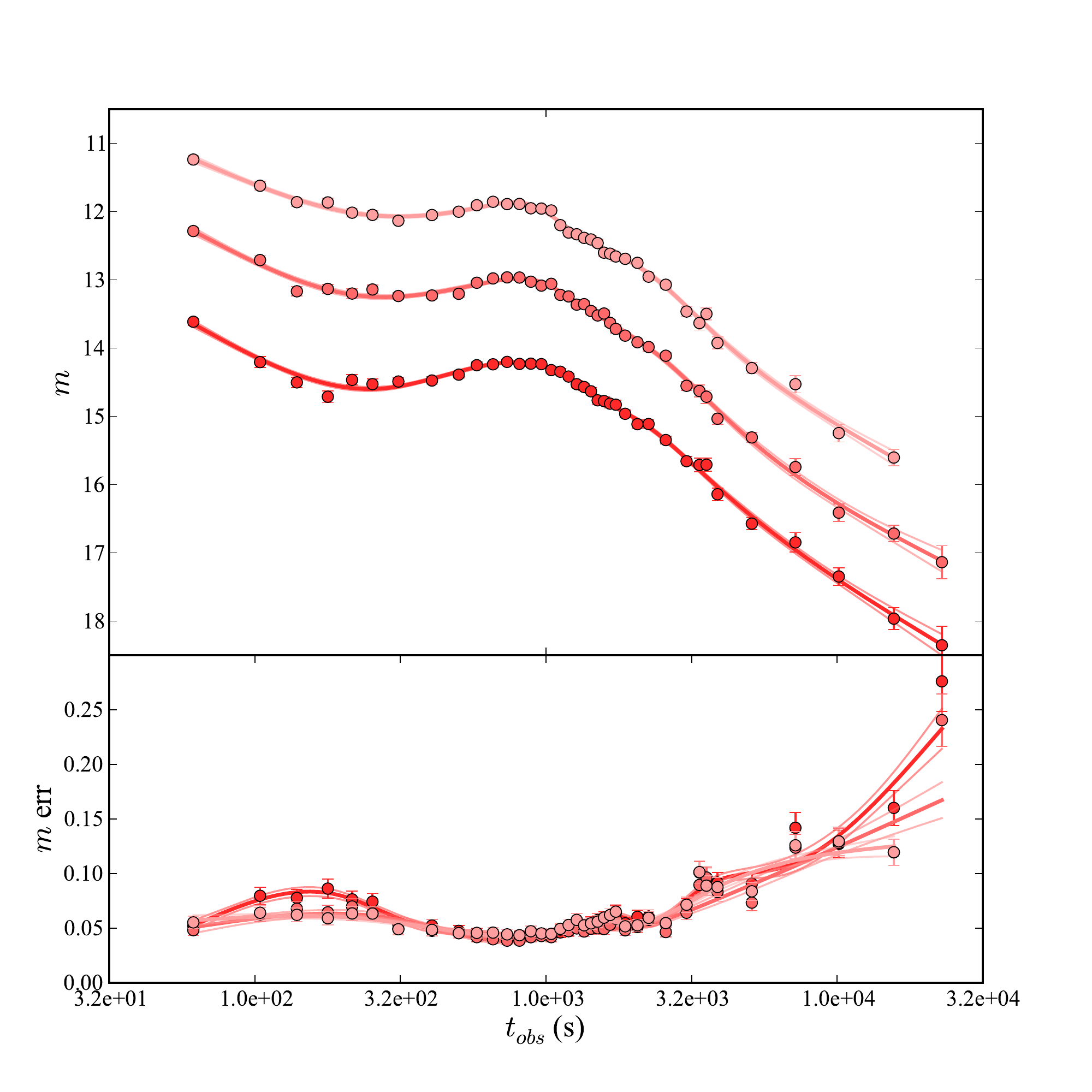}
  \caption{Natural cubic regression spline fits to the PAIRITEL data for use in interpolation.  The upper plot shows the fit to the light curves, found via the GCV criterion to have an optimal number of 8 spline knots for $K_s$, and 14 for each of $J$ and $H$. The bottom plot shows a fit to the photometric uncertainties, assuming each point was uncertain at the $10\%$ level, in order to estimate these values at the interpolation points. The central lines show the optimal model fit, and the lighter outer lines show the model uncertainty. The plot colours are the same as in Figure \ref{fig:lightcurve}.
 }
\label{fig:spline}
\end{figure}

The spline regression provides a model uncertainty for each interpolation point, which must be combined in quadrature with an estimate of what the photometric uncertainty would have been had an observation taken place at that time.  To approximate the latter, we performed another spline fit to the photometric uncertainties for each PAIRITEL filter as a function of time, assuming that each photometric uncertainty (\S \ref{sec:ptel}) itself was uncertain at roughly the $10\%$ level.  The resultant fit is shown in the bottom panel of Figure \ref{fig:spline}; the optimal number of knots was 12, 6, and 7 for $J$, $H$, and $K_s$, respectively.  For each interpolated point, the associated uncertainty at a particular point in time was the mean approximate instrumental uncertainty inferred from this interpolation added in quadrature with the model uncertainty at that time. 

The end result of the interpolation is a series of 4-7 colour SEDs, finely sampled in time.  While some bluer filters are occasionally included in these SEDs, the majority are 5-colour $RIJHK_s$ SEDs from PROMPT and PAIRITEL. While these alone 
cannot strongly break the intrinsic degeneracy between 
$A_V$ and 
$\beta$, the longer lever arm afforded by the more secure late-time SED (\ref{sec:latesed}) gives us strong constraints on both these parameters and the type of dust.  These SEDs can now be used to model the colour change as a function of time.

\subsection{Modelling Colour Change}
\label{sec:colorchange}

\begin{figure*}  
  \begin{minipage}[b]{0.45\linewidth}
  \centering
  \includegraphics[width=9cm]{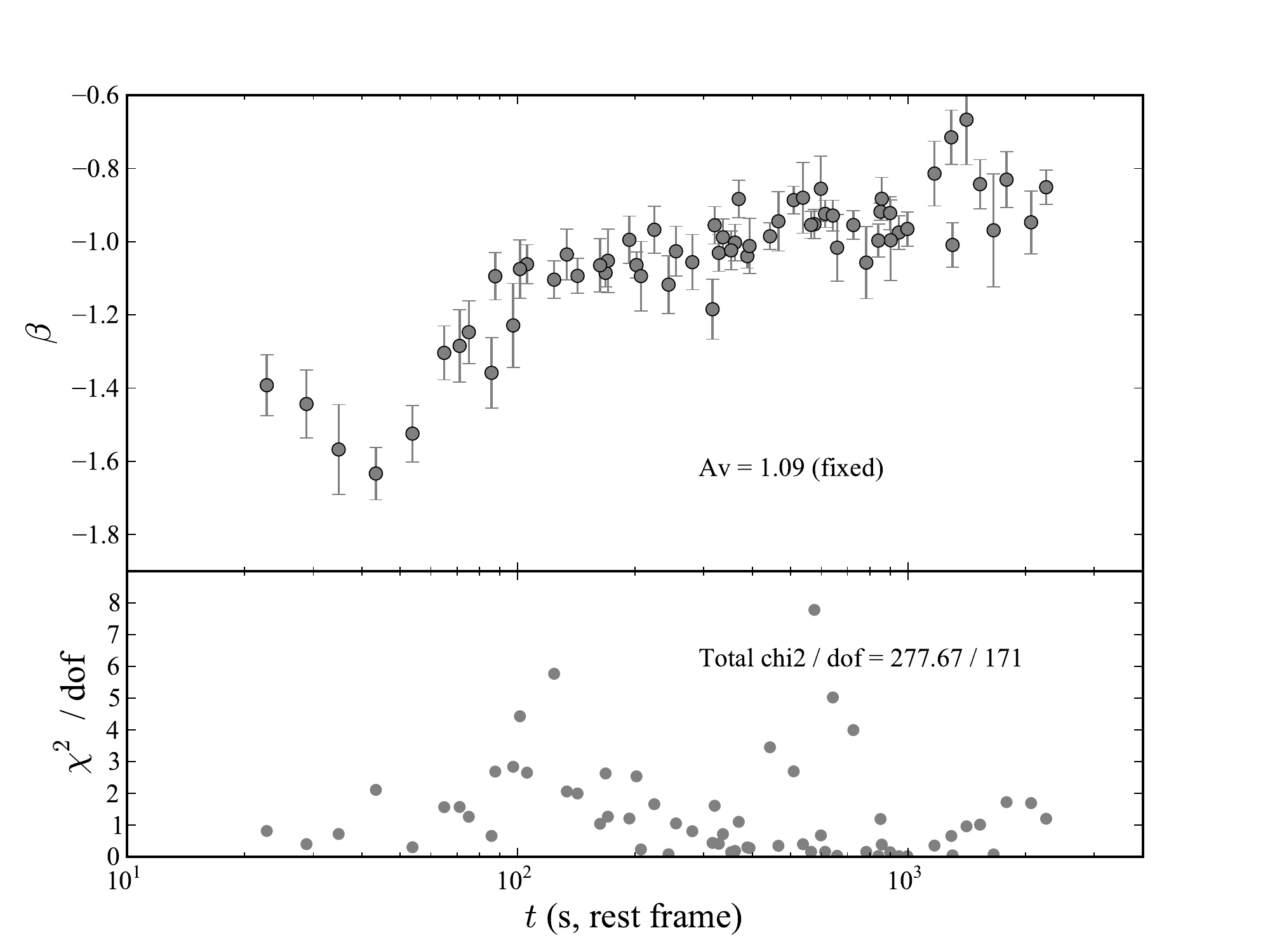}
  \end{minipage}
  \hspace{0.5cm}
  \begin{minipage}[b]{0.45\linewidth}
  \centering
  \includegraphics[width=9cm]{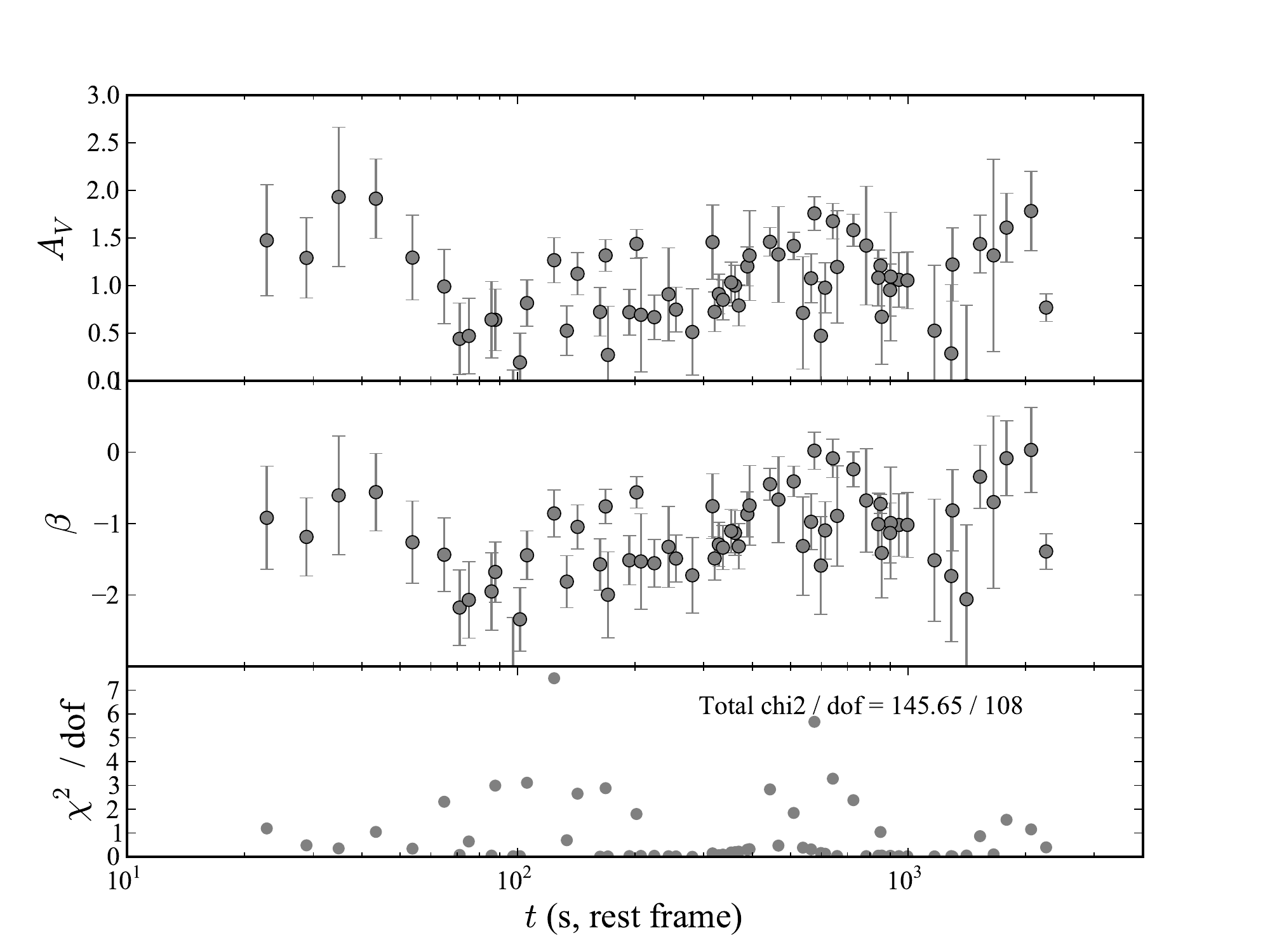}
  \end{minipage}
  \caption{\textit{Left panel:} The best-fit values of $\beta$ for each interpolated SED as a function of time.  The extinction $A_V$ was fixed to the late-time value of 1.09 and FMX-like dust was assumed (see \S\ref{sec:latesed}).  Plotted below is the reduced $\chi^2$ statistic for each individual fit. 
\textit{Right panel:}  Same as the left panel, but where both $A_V$ and $\beta$ were allowed to vary freely.
  }
\label{fig:sedvstime}
\end{figure*}

The simplest way to model the colour change is to assume that the dust properties in the GRB environment are fixed and that only the spectral index $\beta$ is causing the change. Each interpolated SED was fit under this assumption, with $\beta$ left free and the dust parameters fixed at the inferred late-time values from the best-fit dust extinction law (FMX; see \S \ref{sec:latesed}).  The parameter $\beta$ is fit independently at each SED, and no constraints are imposed on how it is allowed to vary with time.  The results are shown in the left panel of Figure \ref{fig:sedvstime}. 

We next explored whether time varying dust extinction signatures could be contributing to the colour change.
The photodestruction of dust is expected to alter both the extinction $A_V$ and selective reddening $R_V$ \citep[e.g.,][]{perna03}.  There is also the possibility that the dust in the local environment of the GRB that would be affected has a significantly different extinction signature than the dust causing the observed late-time absorption.  For simplicity we search for a change in the dust parameters by allowing only $A_V$ to change and keeping all other extinction parameters fixed.

We refit each interpolated SED, allowing both $A_V$ and $\beta$ to vary freely, under no constraints as to how they are allowed to vary as a function of time. The results are shown in 
the right panel of Figure \ref{fig:sedvstime}; they exhibit a drastic change compared with the left panel of this figure in both behavior and overall quality of fit (total $\chi^2$/dof = 146.7/108 vs. $\chi^2$/dof = 277.7/171). The fact that $\beta$ alone is not the dominant source of colour change in this fit gives a preliminary indication that a decrease in absorption may be present.

However, as this simple illustration imposes no constraints regarding how the parameters are allowed to change, the variations in $A_V$ and $\beta$ between epochs are often unphysically large and rapid.  
To account for this, simple, monotonic functional forms for $A_V(t)$ and $\beta(t)$ were assumed:  
\begin{equation}
\label{eqn:av_time}
A_{V}(t) = A_{V,0} + \Delta A_V e^{-t/\tau_{A_V}},
\end{equation}
and
\begin{equation}
\label{eqn:beta_time}
\beta(t) = \beta_{0} + \Delta \beta e^{-t/\tau_{\beta}}.
\end{equation}
In these parameterizations, $A_{V,0}$ and $\beta_0$ represent the late-time values of extinction and spectral index, and $\Delta A_V $ and $\Delta \beta$ correspond to the total change from the early-time values ($t=0$).  No change in the total amount of extinction would thus be consistent with $\Delta A_V =0$.  In these fits, $\Delta A_V $ was required to be positive; that is, we did not allow for an increase in extinction with time.

In fitting this model to the $N$ SEDs constructed from $M$ flux measurements as described in \S \ref{sec:sedvstime}, we fixed the dust extinction model as well the late-time values of $A_V$ and $\beta$ inferred from that model for a variety of the late-time extinction curves shown in Table \ref{tab:extfits}.  In addition to the 4 remaining parameters $\Delta A_V , \Delta \beta, \tau_{A_V},$ and $\tau_\beta$, a flux normalization parameter was fit for each SED, leading to a total of $M-(N+4)$ degrees of freedom. The results are shown in Table \ref{tab:ccfits}. While the precise results are dependent on the assumed late-time extinction model, we note that in all cases, $\Delta A_V $ is found to be positive to a statistically significant level. In other words, under the model assumptions, dust destruction is favored for each extinction law.  The change in spectral index $\beta$, on the other hand, is less favored as the dominant source of colour change, and its behavior with time is more dependent on the assumed extinction law.

\begin{figure}
  \includegraphics[width=9cm]{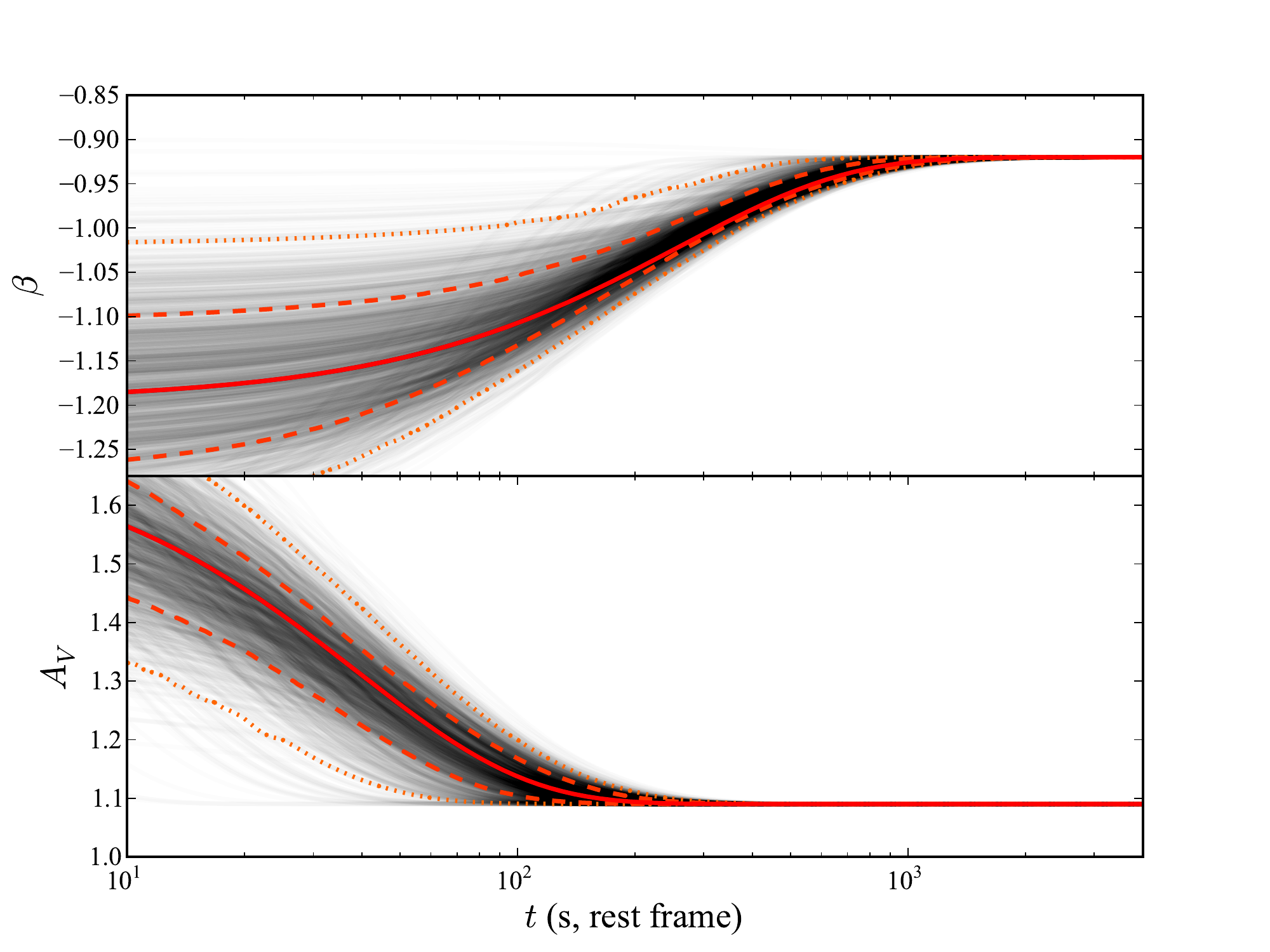}
  \caption{Best-fit values for $A_V(t)$ (mag) and $\beta(t)$ according to the functional forms of Equations \ref{eqn:av_time} and \ref{eqn:beta_time}, simultaneously fit to all available data at each SED interpolation point. FMX-type dust was assumed (\S\ref{sec:latesed}).  
Light-grey lines show the resultant curve from one of the 1500 samples drawn from the multivariate normal distribution from the best-fit model.
The mean value is plotted in red, with confidence contours overplotted with dashed lines (containing 68\% of the curves) and dotted lines (containing 95\% of the curves).
 }
\label{fig:simulfit}
\end{figure}

As with the fits to the late-time dust, the extinction model that yielded the best fit to the colour change was FMX, giving further support for this model over a simple SMC law.  The behaviors of $A_V$ and $\beta$ with time using the best-fit values from this model are presented in Figure \ref{fig:simulfit}. To give a sense of the variance and covariance among the fit parameters, we drew 1500 samples from the resultant best-fit multivariate normal distribution (light-grey curves) with confidence contours overplotted (dashed lines contain 68\% of the curves, dotted 95\%).

\begin{figure}
  \includegraphics[width=9cm]{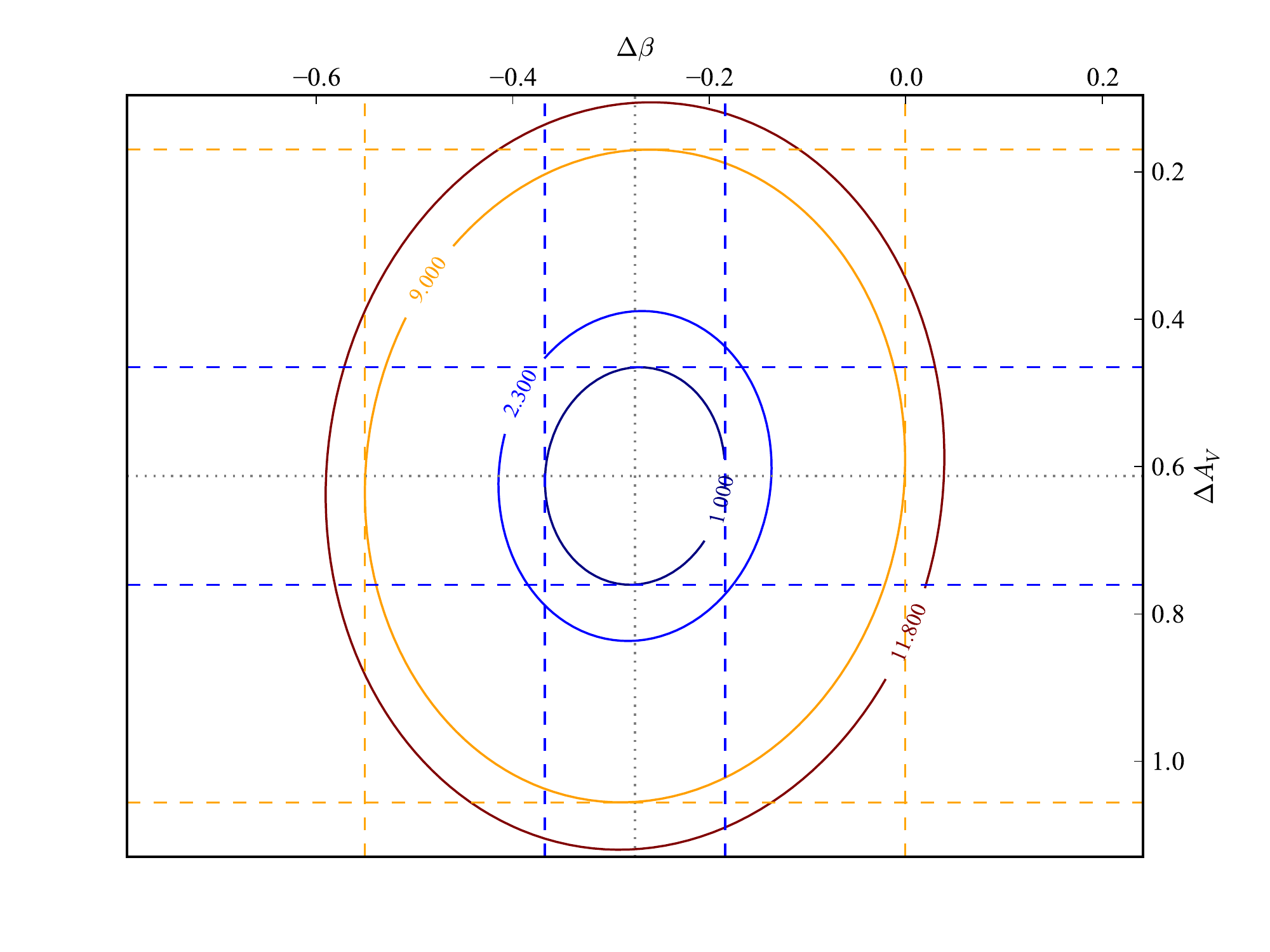}
  \caption{Covariance between parameters $\Delta A_V$ and $\Delta \beta$ from Equations \ref{eqn:av_time} and \ref{eqn:beta_time}. These parameters correspond to the total change in the values of $A_V$ and $\beta$ over the course of the light curve (i.e., a value of $\Delta A_V =0$ would be consistent with no change in the amount of dust). The outer blue ellipse ($\Delta \chi^2=2.3$) and the outermost red ellipse ($\Delta \chi^2=11.8$) show the 68.3\% and 99.73\% confidence intervals for two degrees of freedom.  The dotted blue and orange lines show the projections for a further marginalization to a single parameter of interest, corresponding to the $1\sigma$ and $3\sigma$ confidence intervals for $\Delta A_V$ and $\Delta \beta$ alone.
 }
\label{fig:simulfitmarg}
\end{figure}

Unlike the degeneracy typically seen between $A_V$ and $\beta$ in dust-model fits to optical/NIR SEDs, the 
parameters describing the change in $A_V$ and $\beta$ with time ($\Delta A_V $ and $\Delta \beta $) exhibit only mild covariance.  After marginalizing over all other parameters the covariance is shown in Figure \ref{fig:simulfitmarg}. Subject to all of the model assumptions outlined above, extinction change is expected with about 4$\sigma$ confidence, with a mean value of $\Delta A_V = 0.61 \pm 0.15$ mag.

To guard against the fit falling into a local minimum, we fit each model 1000 times to the data over a randomly selected set of initial conditions for each fit parameter over the following ranges: $0 \le \Delta A_V  \le 2$ mag, $-2 \le \Delta \beta \le$, $0 \le \tau_{A_V} \le 1000$\,s, and $0 \le \tau_\beta \le 1000$\,s.  Each initial normalization parameter value was also randomly selected from a range encompassing all possible physical values.  
The fits given in Table \ref{tab:ccfits} had both the best fits (lowest $\chi^2$) as well as the plurality of convergences for each presented dust model.

\begin{table*}
  \centering
  \begin{minipage}{16.5cm}
  \caption{Results of Color Change Fits}
  \begin{tabular}{llllllll}
  \hline
Dust & $A_{V,0}$ & $\Delta A_{V}$ & $\tau_{A_V}$ & $\beta_{0}$ & $\Delta \beta$ & $\tau_{\beta}$ &  $\chi^2$ $/$ dof \\
Model & (mag) & (mag) & (s) & & & (s) & \\
  \hline
LMC2 & -1.16 (fixed) & $ 0.47 \pm 0.72$ & $ 57.51 \pm 63.09$ & -0.92 (fixed) & $ -0.33 \pm 0.81$ & $ 65.32 \pm 121.31$ & 587.5 / 230 \\
SMC & -0.88 (fixed) & $ 0.60 \pm 0.10$ & $ 44.19 \pm 9.86$ & -0.89 (fixed) & $ -0.28 \pm 0.05$ & $ 513.56 \pm 104.17$ & 383.5 / 230 \\
FMX & -1.09 (fixed) & $ 0.61 \pm 0.15$ & $ 39.00 \pm 13.34$ & -0.92 (fixed) & $ -0.28 \pm 0.09$ & $ 258.90 \pm 71.18$ & 365.6 / 230 \\

  \hline
  \end{tabular}
 
 \label{tab:ccfits}
  \end{minipage}
\end{table*}

\subsection{Optical Rise and Shock Constraints}
\label{sec:earlyrise}
The colour-change modelling described in \S\ref{sec:colorchange} was performed free from assumptions of an underlying light-curve model; each time slice was treated independently and was fit with flux normalization as a free parameter. The analysis above indicates that this colour change is likely caused at least in part by dust destruction (as will be discussed further in \S\ref{sec:discussion}). As such, model-dependent absorption corrections would need to be performed in order to infer the intrinsic light curve, which itself could be produced by multiple emission components, including prompt emission from the central engine.

The second peak at $\sim800$ s (observer frame) is more straightforward to interpret.  The lack of significant colour change across the peak immediately rules out some possibilities for the optical rise.  For instance, the passage of the peak frequency $\nu_m$ of the forward-shock synchrotron emission through the observed bands would produce a temporal rise if it occurred after the initial fireball deceleration. However, this would result in a blue-to-red colour change as $\nu_m$ passes from high to low frequencies; for instance, for expansion into a homogeneous external medium in the slow-cooling regime, the intrinsic spectral index should pass from $\beta \approx \nu^{1/3}$ to $\beta \approx \nu^{-(p-1)/2}$, where $p$ is the electron energy index \citep{sari98}.

The achromatic steep rise and slow decay of the afterglow at this stage shows some of the hallmarks of the onset of the forward shock as the fireball sweeps through the external medium.  If this is the reason for the peak, we can use the time of the peak to obtain an estimate of the initial Lorentz factor $\Gamma.$  Following the methodology of \citet{meszaros06} and \citet{rykoff09a}, and assuming a constant circumburst density profile such as that of the interstellar medium (hereafter ``ISM-like''), we have

\begin{eqnarray}
	\label{eqn:gamma}
	\Gamma_0 \approx 2\Gamma_{\mbox{dec}} & = & 2\left(\frac{3 E_{\rm iso}}{32 \pi n m_{\rm p} c^5 \eta t_{\rm pk,z}^3} \right)^{1/8} \nonumber \\
	& \approx  & 560 \left( \frac{3 E_{\rm iso,52}}{n_{0} \eta_{0.2} t_{\rm pk,z,10}^3} \right)^{1/8}.
\end{eqnarray}

\noindent
In this equation, $E_{\rm iso,52}$ is the  isotropic $\gamma$-ray energy release in units of $10^{52}$\,erg, $n_0$ is the circumburst density in units of cm$^{-3}$, $\eta_{0.2}=\eta/0.2$ is the radiative efficiency, and $t_{\rm pk,z,10}$ is the rest-frame afterglow peak time in units of 10\,s.

This determination requires an estimate of $E_{\rm iso},$ as derived from the spectral model of the gamma-ray data.
For consistency with the analysis of \citet{rykoff09a}, we adopt the value of $E_{\rm iso}$ derived from the methodology of \citet{butler07b}, utilizing a Bayesian analysis with BATSE spectral priors to estimate
$E_{\rm iso}$ using the relatively narrow range of (15--150\,keV) from the BAT. However, the estimate of $\Gamma$ is only weakly dependent on $E_{\rm iso}$ (1/8$^{\rm th}$ power), so the slightly different estimates (by a factor of $< 2$) from {\it Konus-Wind} \citep{golenetskii12a} and {\it Fermi} GBM \citep{gruber12a} does not significantly alter the result.
Using $E_{\rm iso} = 2.1 \times 10^{53}$\,erg and $t_{pk,z} \approx 300$\,s, we find $\Gamma_0 \approx 260 (n_0 \eta_{0.2})^{-1/8}$.

The redshift of GRB\,120119A is comparable to that of GRB\,990123, for which a bright radio flare from the reverse shock was detected. The reverse-shock emission observed for GRB\,990123 peaked at $\sim 1$\,day since burst, at a flux level of $\sim 100$--260\,$\mu$Jy \citep{galama99,kulkarni99} in the GHz range. Our CARMA upper limit on GRB\,120119A (\S\ref{sec:carma}) does not allow us to exclude a radio flare as bright as the one of GRB\,990123. However, GRB\,120119A was also observed with the EVLA beginning 2012 January 21.2 (2.0\,days after the burst) at a mean frequency of 5.8\,GHz \citep{zauderer12a}.  No significant radio emission was detected to a $3\sigma$ upper limit of 34\,$\mu$Jy. This upper limit is below the $\sim 5$\,GHz flux of $164 \pm 100$\,$\mu$Jy measured by \citet{galama99} at $\sim 2$\,day since GRB\,990123. A reverse-shock contribution as bright as the one observed in GRB\,990123 is thus disfavored in the case of GRB\,120119A. 

Hereafter, we check whether the CARMA upper limit can constrain the forward-shock parameters when combined with the optical-to-X-ray observations of GRB\,120119A.
The similarity of the optical/NIR and X-ray temporal slopes, starting from  $\sim 3\times 10^3$\,s since the burst, suggests that the synchrotron frequency $\nu_m$ is below the optical/NIR band \citep[$\nu_m\propto t^{-3/2}$;][]{sari98}, which allows for a determination of an upper limit on $\nu_m$.  A lower limit can be placed on $\nu_m$ by extrapolating the $R$-band optical flux to the time of the CARMA observation with our light-curve model (\S\ref{sec:lc}), under the assumption that the synchrotron self-absorption frequency $\nu_a$ is below the CARMA band at the time of our observation, using
\begin{equation}
F_{\rm 95\,GHz}=F_{\rm opt}\left(\frac{\nu_{\rm opt}}{\nu_m}\right)^{-\beta_{\rm opt}}\left(\frac{\nu_m}{\rm 95\,GHz}\right)^{-1/3}\lesssim 0.99\,{\rm mJy}.
\end{equation}
This gives the bounds $5.49 \times 10^{11} < \nu_m < 4.07 \times 10^{12}$\,Hz and $3.3 \times 10^{2} ~\mu{\rm Jy} < F_m < 1.8 \times 10^{3} ~\mu{\rm Jy}$. In addition, since no strong evidence for a break is observed in the X-ray light curve up to $\sim 1$\,day since the burst, we can constrain the cooling-break frequency to be $\nu_c(1\,{\rm day}) \gtrsim 1$\,keV.

These constraints can be used to make estimates of the circumburst density $n$, the fraction of energy imparted to swept-up electrons $\eta_e$, and the fraction of energy going into magnetic fields $\eta_B$. Using the formalism of \citet{yost03a} and assuming an adiabatically expanding fireball and a constant (ISM-like) circumburst matter density, a solution satisfying our constraints (using $p=2.7$ as estimated above) is 
$n\approx 0.1$\,cm$^{-3}$, $\epsilon_B\approx 5\times10^{-4}$, and $\epsilon_e\approx 0.1$, consistent with values found for other \Swift{} events \citep{liang2008a,melandri10a}. 
We note, however, that this set of parameters would imply a slightly higher flux at the time and frequency of the EVLA observation ($\sim 100$\,$\mu$Jy) than the upper limit of \citet{zauderer12a}, under the assumptions outlined above.

\subsection{Host-Galaxy Properties}
\label{sec:host}

The detection of a nonfading, extended source coincident with the GRB
afterglow days to months after the GRB indicates that we have detected
the GRB host galaxy.   Given the negligible offset and relative
brightness of the source ($<0.2 \arcsec$ and $R=24.8$ mag), we
calculate a very small probability of chance coincidence: $P_{\rm
chance}=4\times10^{-3}$ \citep{Bloom+2002}.

The negligible fading in the $K$ band relative to the late-time data
(and only very marginal fading of the optical counterpart at 1--7
days) indicates that contribution of an afterglow to our photometry of
the host galaxy at 4 days and later is essentially negligible,
allowing us to categorize the properties of the host photometrically.  Our host photometry is given in Table \ref{tab:GRB120119Ahostphot}.
The host colour is observed to be quite red (although not
quite as red as the afterglow), with $R-K=3.6$ mag (2.0 mag AB) after
correction for Galactic reddening. This value is much redder than typical unobscured
GRB hosts \citep{LeFloch+2003,Hjorth+2012} and fairly 
characteristic of the ``dark'' GRB host population (Perley et al., submitted; Malesani et al., in prep.).

\begin{table}
  \centering
  \begin{minipage}{6cm}
  \caption{GRB 120119A Host Photometry}
  \begin{tabular}{lcc}
  \hline
Filter & Magnitude\footnote{Values  in this table have not been corrected for Galactic extinction.} & $1\sigma$ Uncertainty \\
  \hline
$B $ &		$25.50$ & $0.13$ \\
$g'$ &     $25.27$ & $0.05$ \\
$R$ &      $24.60$ & $0.17$ \\
$R$ &      $24.68$ & $0.11$ \\
$I$ &      $23.65$ & $0.13$ \\
$z'$ &     $24.12$ & $0.13$ \\
$K$ &      $20.91$ & $0.13$ \\
$F125W$ &  $23.13$ & $0.04$ \\
$F160W$ &  $23.34$ & $0.07$ \\

  \hline
  \end{tabular}

 \label{tab:GRB120119Ahostphot}
  \end{minipage}
\end{table}

We fit the combined dataset to a range of simple population-synthesis models with \citet{calzetti00}
extinction and constant star-formation history.  Importantly, our models include an empirical treatment 
of nebular emission lines based on the relations of \citet{Kennicutt1998}, \citet{KewleyDopita02}, and \citet{Kewley+2004}.

The resulting SED is nearly flat in $\nu F_\nu$ with no evidence of a Balmer 
break, indicating a young population.  A large upturn is evident in the F125W 
filter relative to all other bands; this feature is highly significant (given 
the excellent quality of the {\it HST} photometry) and cannot be explained by any model 
invoking the stellar continuum alone.  This feature probably results from 
the [O~III] and H$\beta$ emission lines falling within the  F125W 
bandpass --- which, given the strength of the upturn, would imply that these 
features contribute approximately a third of the flux in the F125W broadband 
filter.  This implies an extremely young stellar population.  Based on our SED 
fitting procedure, we estimate a current star-formation rate of $\sim 200$\,$M_\odot\,{\rm yr}^{-1}$ and a stellar mass of only $M \approx 2 \times 
10^{9}\,{\rm M}_\odot$.  The mean attenuation is $A_V \approx 1.8$ mag, indicating a 
quite dusty galaxy.  

These properties suggest a young, 
vigorously star forming, dusty galaxy, with a star-formation rate one hundred 
times larger than that of the Milky Way despite a mass comparable to that of the LMC.  
Given the extreme inferred star-formation rate, we predict that the nebular 
lines of this host should be easily detectable with NIR spectroscopy, and the 
host should be detectable with current radio and millimeter/submillimeter 
facilities as well.  The small size of the galaxy and the concentration of 
most of the light in the core indicates that much of this star formation is 
occurring within a very small volume ($\lesssim 1\arcsec$ in diameter, 
corresponding to 8\,kpc) and over a very short timescale ($\sim 10^7$\,yr) 
without an obvious merger trigger apparent in our imaging, although it is 
possible that even higher-resolution observations (with {\it HST} WFC3-UVIS or ACS, or 
ALMA) will resolve the core into a late-stage merger.      


\section{Discussion and Conclusions}

\label{sec:discussion}

The photodestruction of dust in the  environments of GRBs is expected to occur at early times after the initial blast of high-energy radiation \citep{perna98,waxman00,draine02}.
Simultaneous, multi-colour imaging of the afterglow  (especially during the high-intensity, high-energy emission) is necessary to observe both the decrease in opacity and red-to-blue colour change associated with dust destruction \citep{fruchter01,perna03}.  \citet{perley10} tested for this in the case of GRB 071025 with simultaneous $JHK_s$ and unfiltered optical data (which resembled the $I$ band due to the redness of the afterglow). They binned the optical and NIR measurements to temporally match, and dust models were fit at each four-point SED, allowing both the extinction and intrinsic spectral index $\beta$ to vary. No evidence for a change in absorption was seen, notably during the the first bin, which coincides with the end of the prompt X-ray emission, ruling out dust destruction after the start of the NIR observations ($\sim 150$ s).

We extended upon this methodology for GRB\,120119A by performing interpolation via natural cubic regression splines to the densely sampled PAIRITEL data, which could then be combined with our other photometry (most notably the usually simultaneous PROMPT $R$ and $I$ filters) to create a dense temporal series of early-time SEDs (\S \ref{sec:sedvstime}). The extinction properties inferred from late-time data were held as fixed (Fig. \ref{fig:latesed}), and $\beta$ was fit to each SED. We then explored whether allowing the extinction to vary with time could improve the fit (Fig. \ref{fig:latesed}).

Both $A_V$ and $R_V$ (the latter of which we left fixed in our modelling) are expected to change with time in nontrivial ways under the influence of a high-intensity, high-energy radiation field. Detailed dust-destruction simulations by \citet{perna03} present the time evolution of extinction and reddening for a variety of environments (density and radius), dust distributions, and dust-to-gas ratios.  These simulations highlight the complex interplay between changes in extinction and reddening with time and are highly dependent on the local environment. Note also that these simulations assume a constant luminosity from the X-ray source, and a fixed spectral index ($\beta = 0.5$). A changing luminosity of the photoionizing radiation and a changing spectral index, as are present in GRB prompt emission, further complicate the issue.  Choosing a rigorous parametric model to fit (and inferring meaningful constraints on its parameters) would be very difficult without detailed simulations allowing for the effects of changing luminosities and spectral indices.  

Regardless, our relatively simplistic choice of functional forms for how $A_V(t)$ and $\beta(t)$ could vary (Equations \ref{eqn:av_time} and \ref{eqn:beta_time}) did in fact cause a statistically significant improvement in the overall fit compared to a fixed $A_V$. While not definitively implying dust destruction in GRB\,120119A, simultaneous changes in $A_V$ and $\beta$ can account for the variations in the SED much better than a changing $\beta$ alone.  

Within the assumptions of the model, and by fixing the late-time extinction parameters to the best-fit values derived in \S\ref{sec:latesed}, the extinction $A_V$ is expected to decrease by an average of $0.61 \pm 0.15$ mag
(Fig. \ref{fig:simulfit}), with at least $\sim 0.15$ mag of extinction change to a 99.73\% level ($3\sigma$; Fig. \ref{fig:simulfitmarg}).  Alternative parametric models may be more appropriate, but the limited spectral coverage at each temporal point prevents stringent model selection. Moreover, the inference of dust destruction appears to be robust against the exact choice of assumed late-time extinction curve, as all reasonable tested dust models resulted in this conclusion (after fixing the late-time $A_V$ and $\beta$ to their best-fit values for each particular model).  In further support of the general conclusion of dust destruction, we note that the bulk of the colour change is ongoing during the tail end of the high-intensity X-ray emission, during which dust destruction is expected to occur \citep{fruchter01}.

Finally, we note that this work highlights the importance of well-sampled, multi-colour early-time observations afforded by small and nimble fully robotic telescopes quickly training on the GRB location.  Delays of seconds matter at this stage of the evolution of the GRB, and the importance of observation speed outweighs that of observation depth.  The combination of shallow-and-fast robotic follow-up observations with deep late-time observations across the electromagnetic spectrum is necessary to illustrate the full story behind the evolution of GRB afterglows,
which still to this day are revealing new insights.

\begin{table*}
  \centering
  \begin{minipage}{12cm}
  \caption{Photometry of GRB\,120119A}
  \begin{tabular}{lcccrc}
  \hline
Instrument & Filter & $t_{\rm mid}$\footnote{Time since the \Swift{} trigger of the midpoint of the exposure.} & Exposure Time & 
Magnitude\footnote{Magnitudes in this table have not been corrected for Galactic extinction
  ($E(B-V) = 0.093$\,mag).  Observations in the
  $g^{\prime}$-, $r^{\prime}$-, $i^{\prime}$-, and $z^{\prime}$-bands are reported on 
  the AB magnitude system.  $B$-, $V$-, $R$-, $I$-, $J$-, $H$-, and $K$-band
  observations are referenced to Vega.} & Mag. Uncertainty \\
(UT) & & (s) & (s) & & \\
  \hline

KAIT   &  clear & $299.00$   & $20.0$  &  $16.96$ &  $0.07$  \\
KAIT   &  clear & $399.00$   & $20.0$  &  $17.02$ &  $0.09$  \\
KAIT   &  clear & $498.00$   & $20.0$  &  $16.87$ &  $0.06$  \\
KAIT   &  clear & $598.00$   & $20.0$  &  $16.76$ &  $0.10$  \\
KAIT   &  clear & $698.00$   & $20.0$  &  $16.75$ &  $0.06$  \\
KAIT   &  clear & $796.00$   & $20.0$  &  $16.75$ &  $0.07$  \\
KAIT   &  clear & $896.00$   & $20.0$  &  $16.70$ &  $0.07$  \\
KAIT   &  clear & $995.00$   & $20.0$  &  $16.68$ &  $0.07$  \\
KAIT   &  clear & $1095.00$  & $20.0$  &  $16.95$ &  $0.06$  \\
KAIT   &  clear & $1195.00$  & $20.0$  &  $17.04$ &  $0.06$  \\

  \hline
  \end{tabular}
 
  This table is published in its entirety in the electronic edition of MNRAS. A portion is shown here for guidance regarding its form and content.

 \label{tab:GRB120119Aphot}
  \end{minipage}
\end{table*}

\section*{Acknowledgments}

PAIRITEL is operated by the Smithsonian Astrophysical Observatory (SAO) and was made
possible by a grant from the Harvard University Milton Fund, a camera loan from the University
of Virginia, and continued support of the SAO and UC Berkeley. The PAIRITEL project is further
supported by NASA/{\it Swift} Guest Investigator grants NNX10AI21G and NNX12AD73G.
We utilized observations obtained at the Gemini Observatory, which is operated by the 
Association of Universities for Research in Astronomy, Inc., under a cooperative agreement 
with the NSF on behalf of the Gemini partnership: the NSF (United 
States), the Science and Technology Facilities Council (United Kingdom), the 
National Research Council (Canada), CONICYT (Chile), the Australian Research Council (Australia), 
Minist\`{e}rio da Ci\^{e}ncia, Tecnologia e Inova\c{c}\~{a}o (Brazil), 
and Ministerio de Ciencia, Tecnolog\'{i}a e Innovaci\'{o}n Productiva (Argentina).
The Liverpool Telescope is operated on the island of La Palma by Liverpool John Moores University in the Spanish Observatorio del Roque de los Muchachos of the Instituto de Astrofisica de Canarias with financial support from the UK Science and Technology Facilities Council.
KAIT and its ongoing operation were made possible by donations from Sun
Microsystems, Inc., the Hewlett-Packard Company, AutoScope
Corporation, Lick Observatory, the NSF, the University of California,
the Sylvia and Jim Katzman Foundation, and the TABASGO Foundation. 
Some of the data presented herein were obtained at the W. M. Keck Observatory, which is operated as a scientific partnership among the California Institute of Technology, the University of California, and NASA; the Observatory was made possible by the generous financial support of the W. M. Keck Foundation. We wish to extend special thanks to those of Hawaiian ancestry on whose sacred mountain we are privileged to be guests. We are grateful to the staffs at the Gemini, Keck, and Lick Observatories for their assistance.
This publication has made use of data obtained from the \Swift{} interface
of the High-Energy Astrophysics Archive (HEASARC), provided by
NASA's Goddard Space Flight Center. We sincerely thank the \Swift{} team for the rapid public dissemination, calibration, and analysis of the \Swift{} data.  We also thank Scott Barthelmy for his invaluable efforts in creating and maintaining the GCN system.

C.G.M. acknowledges support from the Royal Society.
A.V.F.Õs group at UC Berkeley has received generous financial assistance from Gary and Cynthia Bengier, the Christopher R. Redlich Fund, the Richard and Rhoda Goldman Fund, the TABASGO Foundation, NSF grant AST-1211916, and NASA/{\it Swift} grants NNX10AI21G and NNX12AD73G.
A.G. acknowledges funding from the Slovenian Research Agency and from the Centre of Excellence for Space Science and Technologies SPACE-SI, an operation partly financed by the European Union, the European Regional Development Fund, and the Republic of Slovenia.
Support for this work was provided by NASA to D.A.P. through Hubble Fellowship grant HST-HF-51296.01-A awarded by the Space Telescope Science Institute, which is operated for NASA by AURA, Inc., under contract NAS 5-26555. We
dedicate this paper to the memory of Weidong Li, who developed the
software for KAIT automatic follow-up observations of GRBs;
we deeply miss his friendship and collaboration, which
were tragically taken away from us much too early.


\bibliography{120119A}{}
\bibliographystyle{mn2e}

\label{lastpage}

\end{document}